\documentclass[aps,prd,preprint,groupedaddress,showpacs]{revtex4-1}	
        
\usepackage{graphicx}	 		
\usepackage{amsmath,amsfonts,amsbsy}
\usepackage{slashed}

\usepackage[scr=boondoxo,scrscaled=1.05]{mathalfa}

\usepackage{hyperref}

\newcommand{\overbar}[1]{\mkern 1.5mu\overline{\mkern-1.5mu#1\mkern-1.5mu}\mkern 1.5mu}

\newcommand{\laser}{\mathcal{A}}
\newcommand{\laserforslash}{\mathcal{A}\,\,}
\newcommand{\ee}{\mathrm{e}}
\newcommand{\ecc}{\tau}

\newcommand{\psl}{\slashed{p}}

\newcommand{\ksl}{\slashed{k}}

\newcommand{\lasersl}{\slashed{\laserforslash}\!\!} 
\newcommand{\ssl}{\slashed{s}}

\newcommand{\lasers}{\laser^{*}}

\newcommand{\laserssl}{\slashed{\laser^*}}

\newcommand{\ve}{\varepsilon}

\newcommand{\psiV}{\psi_{_{\mathrm{V}}}}

\newcommand{\intfkD}{\int{\bar{}\kern-0.45em d}^{\,D\!}k\,}
\newcommand{\intfsD}{\int{\bar{}\kern-0.45em d}^{\,D\!}s\,}
\newcommand{\intfsDuv}{\int_{_\mathrm{UV}}\frac{d^{D}s}{(2\pi)^D}\,}
\newcommand{\intftD}{\int{\bar{}\kern-0.45em d}^{\,D\!}t\,}
\newcommand{\kdotx}{k{\cdot}x}

\newcommand{\intfps}{\int{\bar{}\kern-0.45em d}^{\,3}p\,}
\newcommand{\intps}{\int\frac{{\bar{}\kern-0.45em d}^{\,3}p\,}{2E^*_p}}
\newcommand{\intfp}{\int{\bar{}\kern-0.45em d}^{\,4}p\,}
\newcommand{\intfqs}{\int{\bar{}\kern-0.45em d}^{\,3}q\,}
\newcommand{\intfpqs}{\int{\bar{}\kern-0.45em d}^{\,3}p\,\,\,{\bar{}\kern-0.45em d}^{\,3}q\,}

\newcommand{\Ab}{\mathrm{A}}
\newcommand{\Em}{\mathrm{E}}
\newcommand{\Prop}[1]{\mathrm{P}_{#1}}
\newcommand{\PropB}[1]{\overbar{\mathrm{P}}_{#1}}
\newcommand{\PropI}[1]{\mathrm{P}^{-1}_{#1}}
\newcommand{\In}{\mathrm{I}}
\newcommand{\Out}{\mathrm{O}}
\newcommand{\duv}{\delta_{_{^\mathrm{UV}}}}

\newcommand{\kx}{k{\cdot}x} 
\newcommand{\pk}{p{\cdot}k}

\newcommand{\plaser}{p{\cdot}\laser}
\newcommand{\plasers}{p{\cdot}\lasers}
\newcommand{\laserv}{\laser{\cdot}\laser}
\newcommand{\laserm}{\lasers\!{\cdot}\laser }
\newcommand{\laservs}{\lasers\!{\cdot}\lasers}

\newcommand{\mstar}{\mathscr{M}}

\newcommand{\mstarsl}{\slashed{\mstar}}

\newcommand{\J}{J^{\ecc}}
\newcommand{\Js}{J^{\ecc *}}

\newcommand{\psiVbare}{\psiV^{^{_\mathrm{B}}}}
\newcommand{\mbare}{m^{^{_\mathrm{B}}}}
\newcommand{\mstarbare}{\mstar_\mu^{^{_\mathrm{B}}}}
\newcommand{\mstarbaremin}{\mstar^{^{_\mathrm{B}}}}
 
\newcommand{\psiVphys}{\psiV}
\newcommand{\mphys}{m}
\newcommand{\mstarphys}{\mstar_\mu}
\newcommand{\deltamstar}{\delta_{_{^{\!\!\mstar}}}}
\newcommand{\sigmasl}{\slashed{\Sigma}_{_{\!\!\mstar}}}

\graphicspath{{./pictures/}}

\begin{document}

    
\title{Renormalisation of the Volkov propagator}

\author{Martin~Lavelle and David~McMullan}
\affiliation{Centre for Mathematical Sciences\\University of Plymouth \\
Plymouth, PL4 8AA, UK}

\date{\today}


\begin{abstract}
The perturbative description of an electron propagating in a plane wave background is developed and loop corrections analysed. The ultraviolet divergences and associated renormalisation are studied using the sideband framework within which the  multiplicative form of the corrections becomes manifest. An additional renormalisation beyond that usually expected is identified and interpreted as a loop correction to the background induced mass term. Results for the strong field sector are conjectured. 
\end{abstract}
 
\maketitle

\section{Introduction} 
The Volkov solution~\cite{Volkov:1935zz} for an electron in a  plane wave background is one of the key theoretical  building blocks underpinning our understanding of how matter interacts with a laser. As  quantum effects become significant, strong field  techniques from quantum electrodynamics, QED, are required.  Understanding potential new physics in this high intensity regime is of clear importance and, in turn,  should influence plans for future facilities and experiments.  

The Volkov solution has been extensively studied over the years and applied to a wide class of problems in both linearly and circularly polarised backgrounds, see for example~\cite{Reiss:1966A}\cite{Brown:1964zz}\cite{Nikishov:1964zza}\cite{Neville:1971uc}\cite{Dittrich:1973rn}\cite{Dittrich:1973rm}\cite{Kibble:1975vz}\cite{Mitter:1974yg}\cite{Ritus:1985review}\cite{Ilderton:2012qe}\cite{Lavelle:2013wx}\cite{Lavelle:2014mka}\cite{Lavelle:2015jxa}\cite{King:2014wfa}. Working in the full elliptic class of polarisations  allows for a much clearer description of these systems and helps clarify some of their   physical content~\cite{Lavelle:2017dzx}. In particular, this more general approach  shows that  the laser induced mass shift is actually independent of the eccentricity of the background.   
  
Loop corrections in a laser background have been looked at several times, as for example in~\cite{Brouder:2002fz}\cite{ Heinzl:2011ur}\cite{DiPiazza:2011tq}\cite{Meuren:2011hv}\cite{Podszus:2018hnz}\cite{Ilderton:2019kqp}. Unitarity{} arguments are often used to   directly link loop corrections to effective  cross-sections. It has been argued, see for example~\cite{Baier:1975ys},  that the laser background has no impact on the renormalisation of the the{}ory. To have confidence in this result, it is important to probe the loop structures and associated renormalisation of the theory in a variety of ways. In this paper we will do this by taking a weak field perspective which has the advantage that standard perturbative techniques can be directly applied. 

The propagation of an electron in a laser background is often denoted by a double line. This notation represents the inclusion of multiple interactions with the laser. A physical way to think of this is that the double line incorporates all degenerate processes, i.e., the emission and absorption of photons indistinguishable from the background. This is reminiscent of the Lee-Nauenberg approach to the infrared problem~\cite{Lee:1964is},  see also \cite{Lavelle:2005bt}.  

\begin{figure}[htb]
	\[
	\raisebox{-0.5cm}{\includegraphics{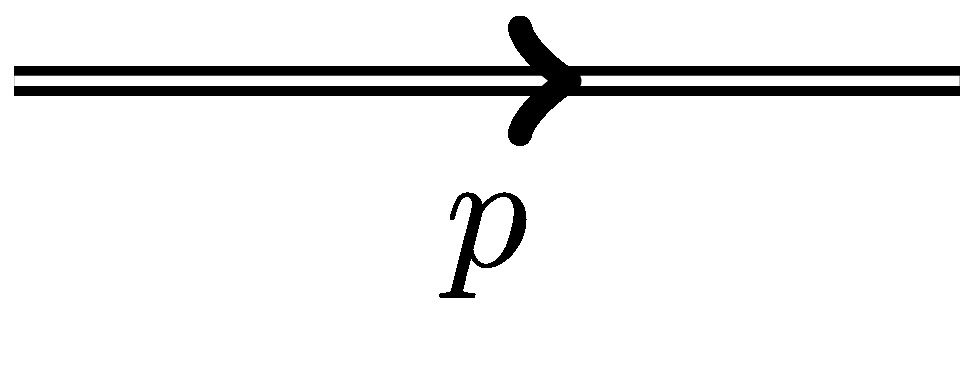}}\ =\sum_{\substack{\mathrm{laser}\\ \mathrm{interactions}}}\raisebox{-0.8cm}{\includegraphics{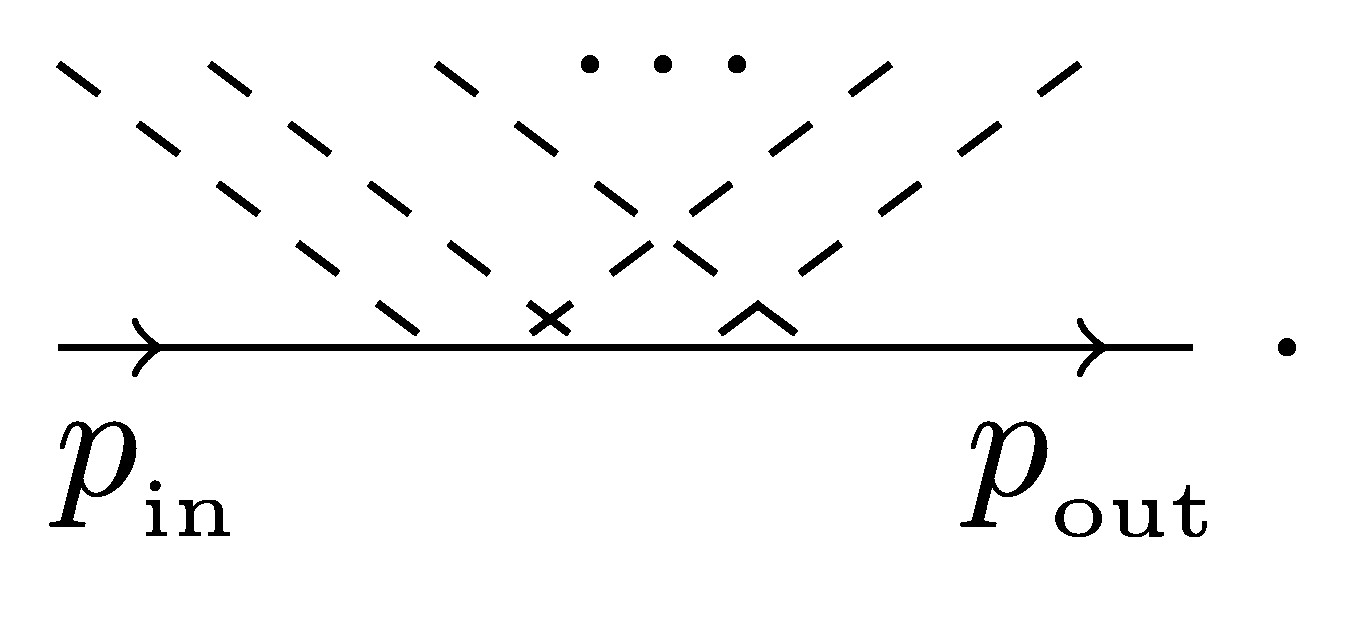}}
	\]
	\caption{Double line representation of an electron propagating in a background at tree level.}
	\label{fig:double_line}
\end{figure}

\begin{figure}[htb]
	\[
\raisebox{-0.6cm}{\includegraphics{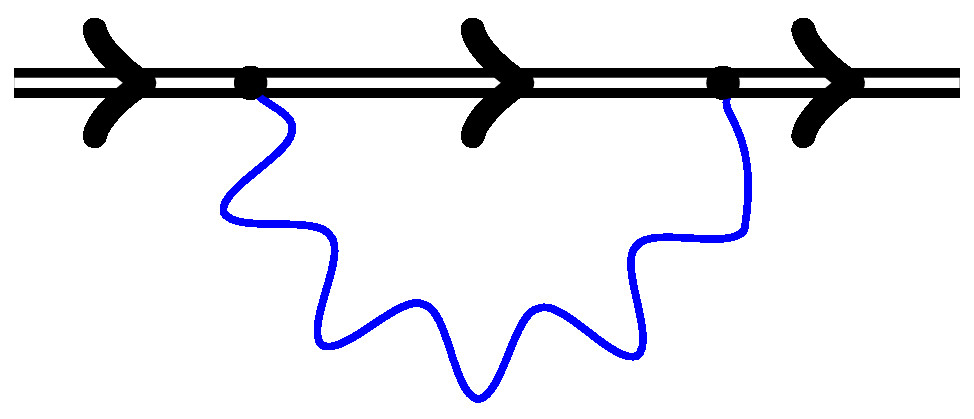}}\ =\sum_{\substack{\mathrm{laser}\\ \mathrm{interactions}\\ \mathrm{and\; loops}}}\raisebox{-0.8cm}{\includegraphics{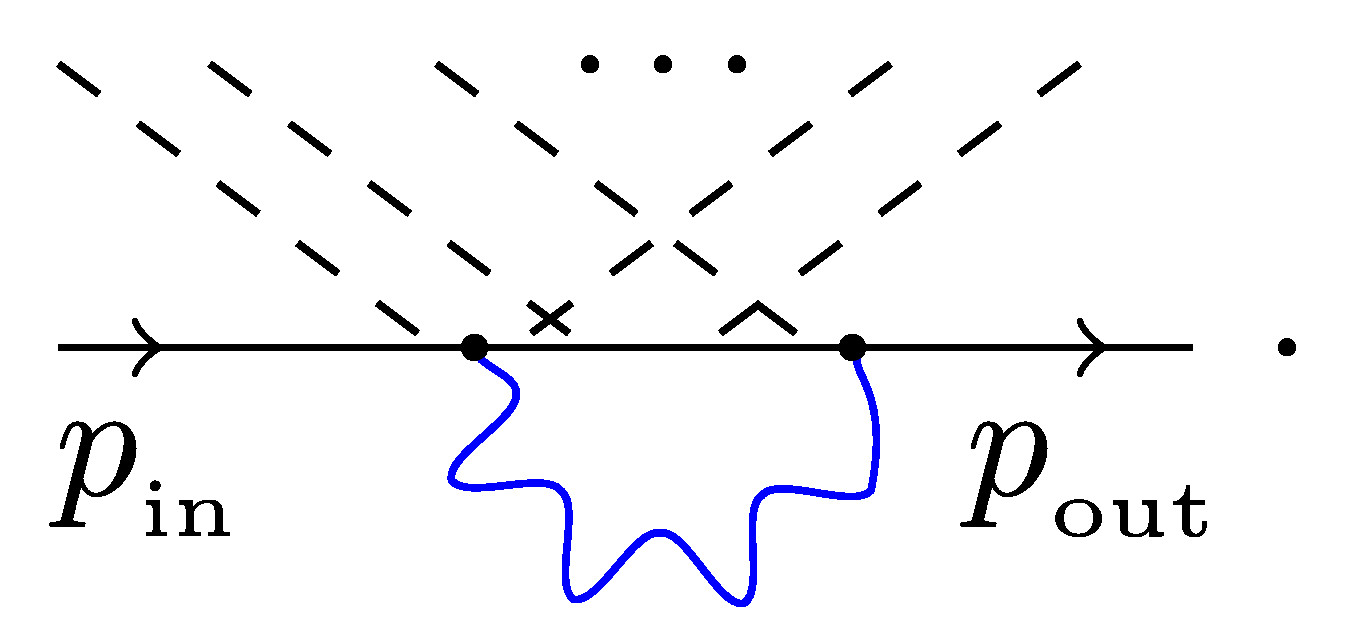}}	
\]
\caption{One loop correction to the double line representation.}
	\label{fig:double_line_loop}
\end{figure}

We take the double line to mean the two point function for the Volkov field in the plane wave background, see Fig.~\ref{fig:double_line}. 
We will, in this paper, make  this link precise in terms of emission into and absorption from the laser. Throughout this paper, we will distinguish between absorption (dashed lines coming in from the left) and emission (dashed lines going out to the right). 
The incorporation of loops in the weak field limit will then follow using standard perturbative methods.
This will then allow us to better understand the way in which loops are added to the double line, see Fig.~\ref{fig:double_line_loop}. 
As we shall see, clarifying the links between these descriptions of matter propagating in a laser will reveal important points about the  renormalisation of such charges.

In this paper we will  study the renormalisation of the theory describing an electron propagating in a plane wave background. This analysis will start in the weak intensity regime and there we will calculate the ultraviolet divergences that arise at one loop. Our loop calculations will be in the Feynman gauge. Polarisation effects  will be clarified through working with the full elliptic class at all times. As well as the naively expected ultraviolet structures, that are independent of the background, we will identify an additional correction to the laser induced mass shift. We show through explicit calculations of  higher order laser interactions that they are renormalised by the same additional  correction, and conjecture that this is universal for this class of backgrounds. Renormalisation is most easily studied within a momentum space description of the theory, so we conclude with a discussion on how a consistent momentum space language can be applied to this system where translation invariance has been broken by the  background field, and conjecture all orders results.

\section{The perturbative set up}
An electron propagating through a laser can absorb  multiple photons from  the background. Additionally, such an electron can emit photons which are degenerate with, and indistinguishable from, the background.
Both types of interactions are, as we shall discuss, required for the double line  description. If, however, the electron emits a photon that is distinguishable from the background then this corresponds to non-linear Compton scattering rather than propagation. 

We will argue  that summing in a suitable way over all such degeneracies  leads to the Volkov description~\cite{Volkov:1935zz}  of an electron propagating through such a background. What is more, this will allow for a direct route to the incorporation of loop corrections in such processes and hence the  renormalisation of the theory.

The  momentum of an electron in a plane wave background can  be decomposed into some initial momentum $p$, along with multiples of the null momentum $k$ characterising the background.  We denote by $\Prop{n}$ the resulting propagator after $n$ net laser absorptions:
\begin{equation}\label{eq:Prop}
  \Prop{n}=\frac{i}{\psl+n\ksl-m+i\epsilon}\,.
\end{equation}
Note that in terms of the overall momentum for the electron, we view an emission as a negative absorption from the laser. So if there were two absorptions and one emission, say, then $n=1$.
This compact notation for the propagators will provide the building blocks for our description of both tree and loop corrected propagation. 

\begin{figure}[htb]
	\[
\raisebox{-0.6cm}{\includegraphics{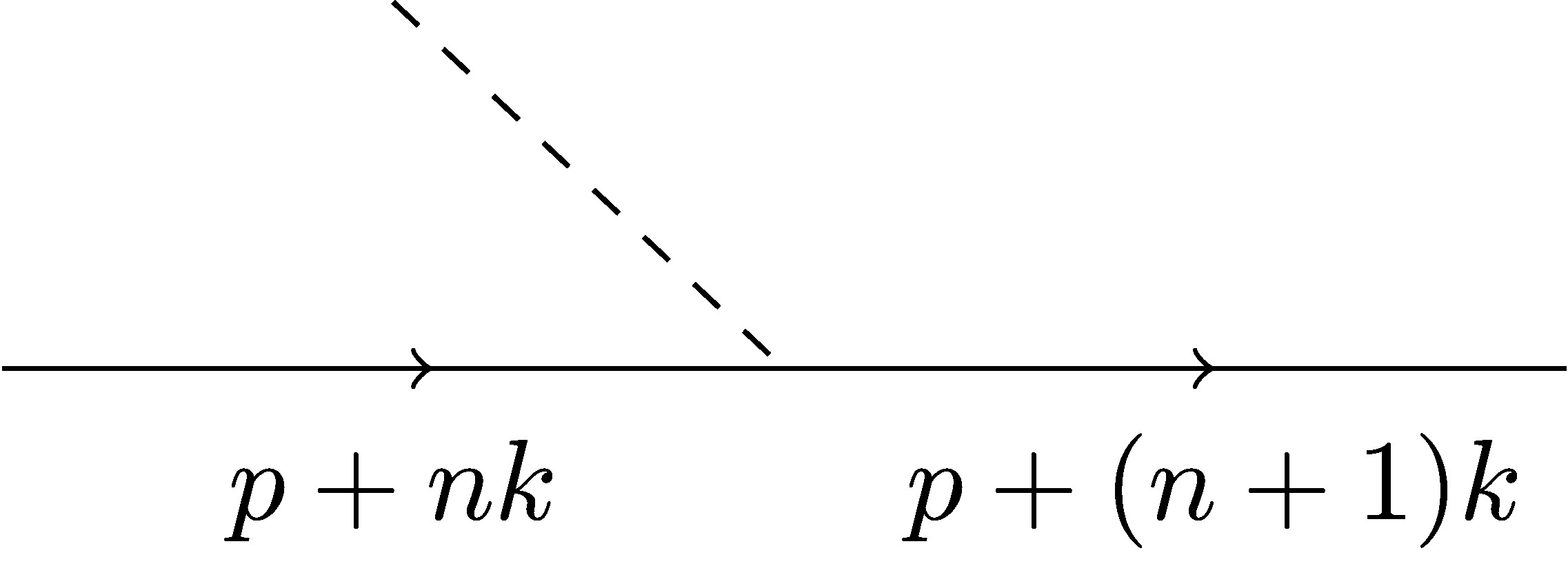}}
=\Prop{n+1}\Ab\Prop{n}
	\]
	\caption{Single absorption from the background.}
	\label{Fig:Ab}
\end{figure}

For example, an additional absorption by the electron is  described by the incoming interaction shown in Fig.~\ref{Fig:Ab}, 
where the absorption factor $\Ab$ between the propagators is given by
\begin{equation}\label{eq:Ais}
	\Ab=-i\lasersl\,.
\end{equation}
Here $\laser_\mu$ is essentially the coupling, $e$, times the Fourier component of the  classical potential for an elliptically polarised plane wave. We will expand on this terminology later, but see also~\cite{Lavelle:2017dzx} for more details on this formalism and the  connection to the Stokes' parameter description of the background.

\begin{figure}[htb]
	\[
 \raisebox{-0.6cm}{\includegraphics{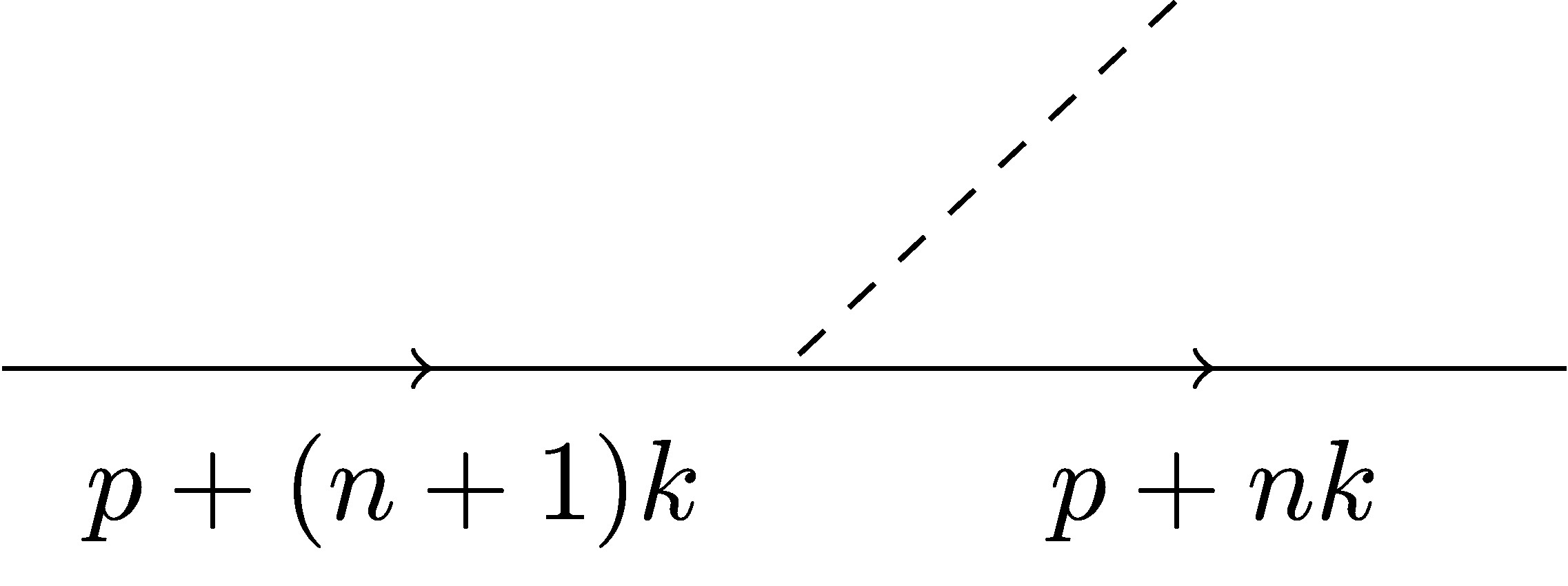}}
=\Prop{n}\Em\Prop{n+1}
	\]
	\caption{Single emission into the background.}
	\label{Fig:Em}
\end{figure}

The associated emission of a photon degenerate with the background is described by the outgoing process given in Fig.~\ref{Fig:Em},
where  the emission factor $\Em$ is given by 
\begin{equation}\label{eq:Eis}
	\Em=-i\laserssl\,.
\end{equation}

It is helpful here to clarify the notation being used. By $\laserssl$ we mean the slashed version of the conjugated field, so  $\lasersl^*=\lasers_\mu\gamma^\mu$. This is a useful shorthand for the unambiguous expression for the dual field
\begin{equation}\label{eq:duality}
  \overbar{\lasersl}:=\gamma_0\lasersl^\dagger\gamma_0\equiv\lasers_\mu\gamma^\mu\,.
\end{equation} 
Note that acting on the propagators we have the duality relation $\PropB{n}= -\Prop{n}$ and on the absorption term $\overbar{\Ab}=-\Em$. 
The duality transformation  needs to respect the time-ordering implicit in the $i\epsilon$ prescription. This means that formally we should take $\overbar{\epsilon}=-\epsilon$. 
Overall, the processes in Figs.~\ref{Fig:Ab} and~\ref{Fig:Em} are (anti) dual to each other in the sense that
\begin{equation}\label{eq:dualitytree}
  \overbar{\Prop{n+1}\Ab\Prop{n}}=-\Prop{n}\Em\Prop{n+1}\,.
\end{equation} 
  
\begin{figure}[htb]
	\[
\raisebox{-0.6cm}{\includegraphics{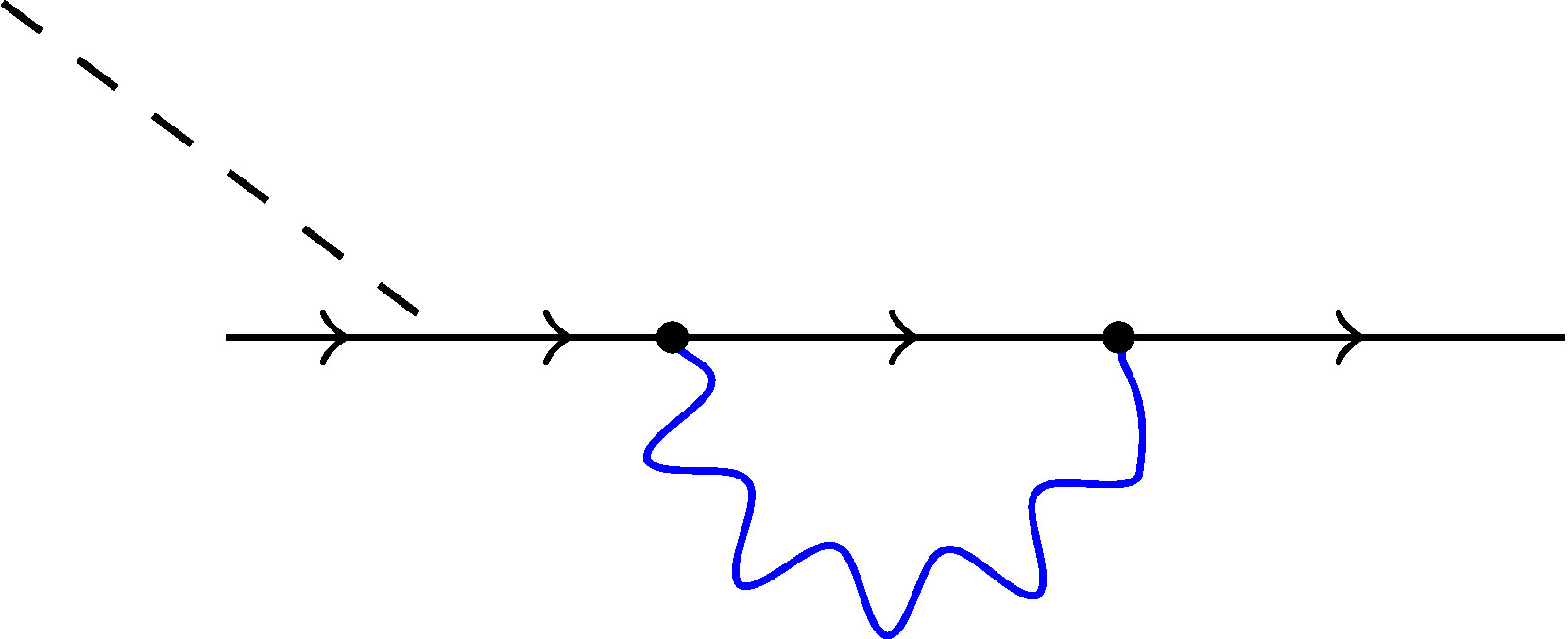}} \ +\ 
\raisebox{-0.6cm}{\includegraphics{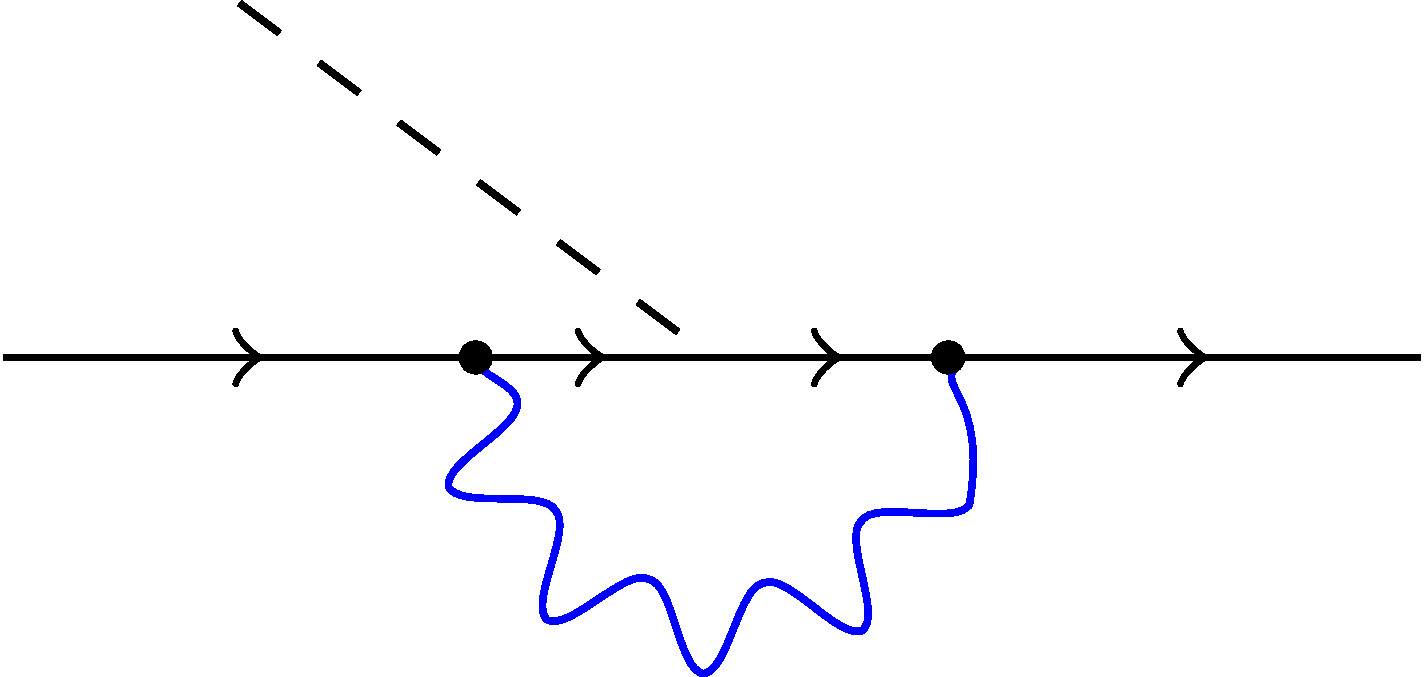}} \ +\ 
\raisebox{-0.6cm}{\includegraphics{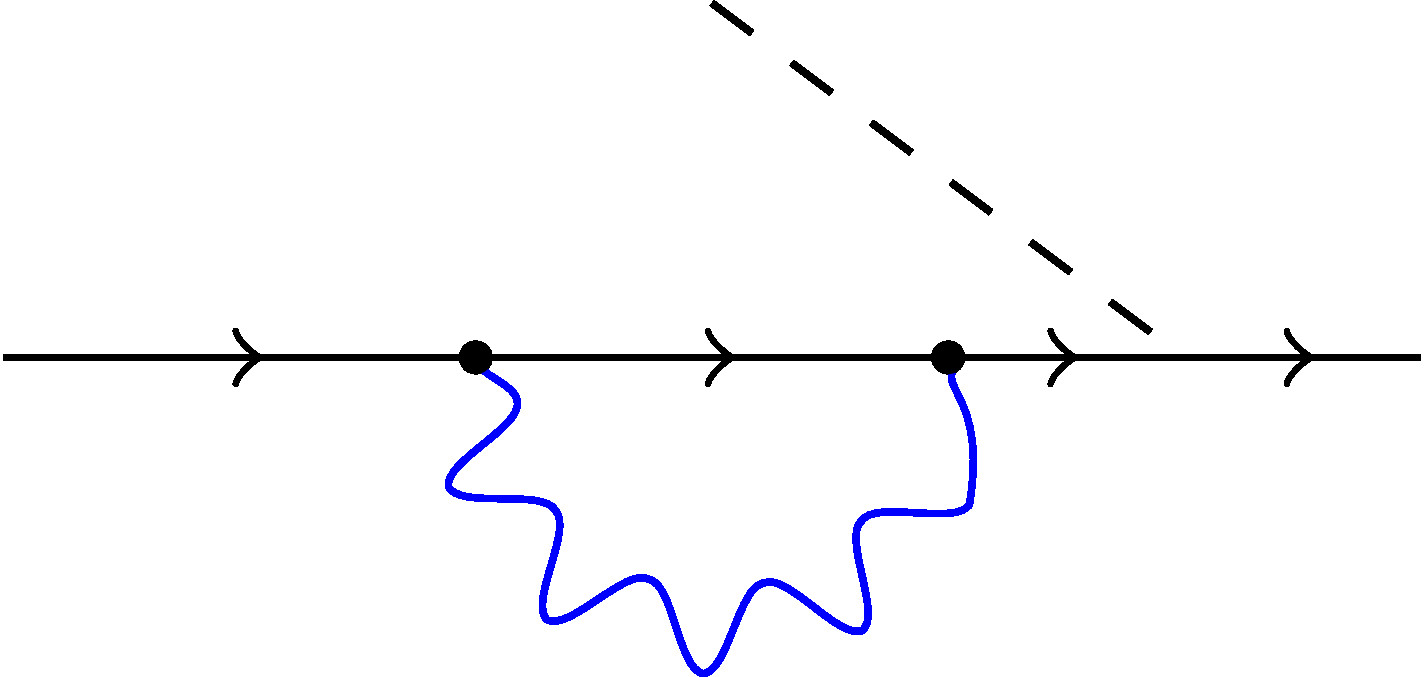}}
	\]
	\caption{Single absorption with a loop correction.}
	\label{Fig:In}
\end{figure}

\begin{figure}[htb]
	\[
\raisebox{-0.6cm}{\includegraphics{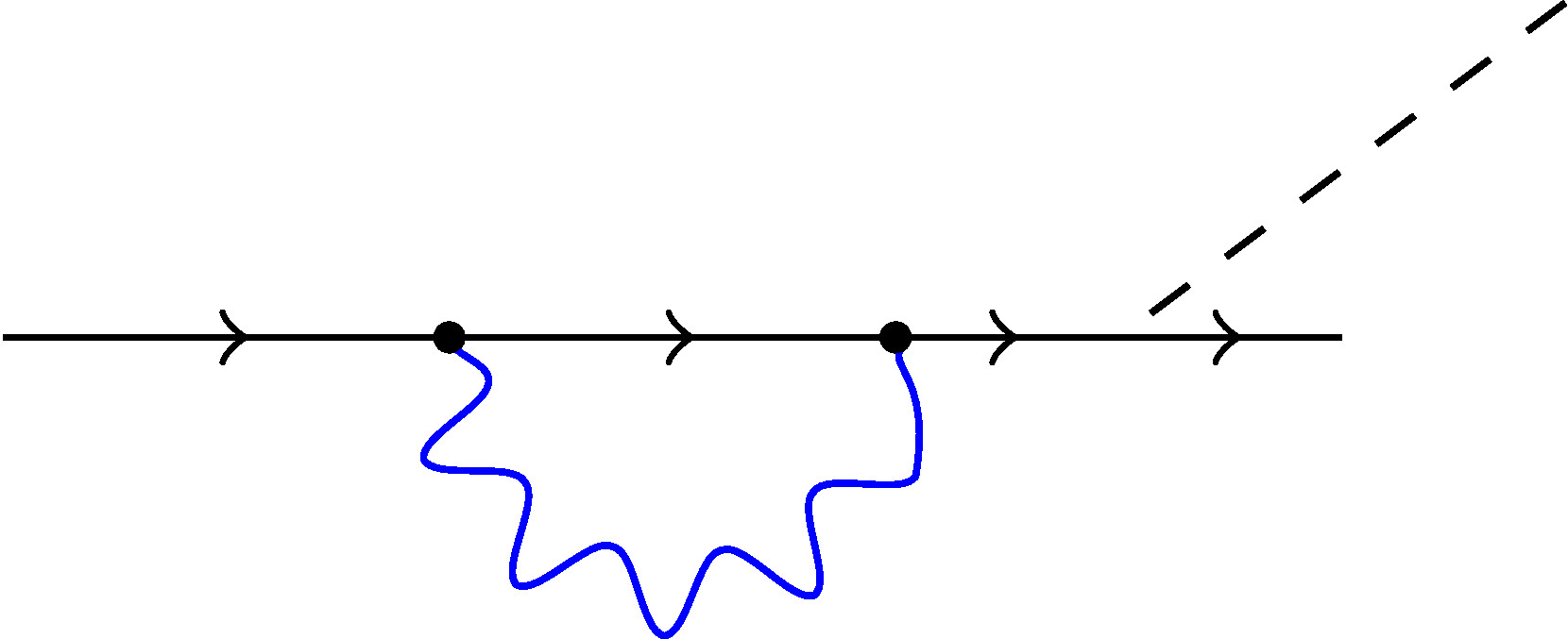}} \ +\ 
\raisebox{-0.6cm}{\includegraphics{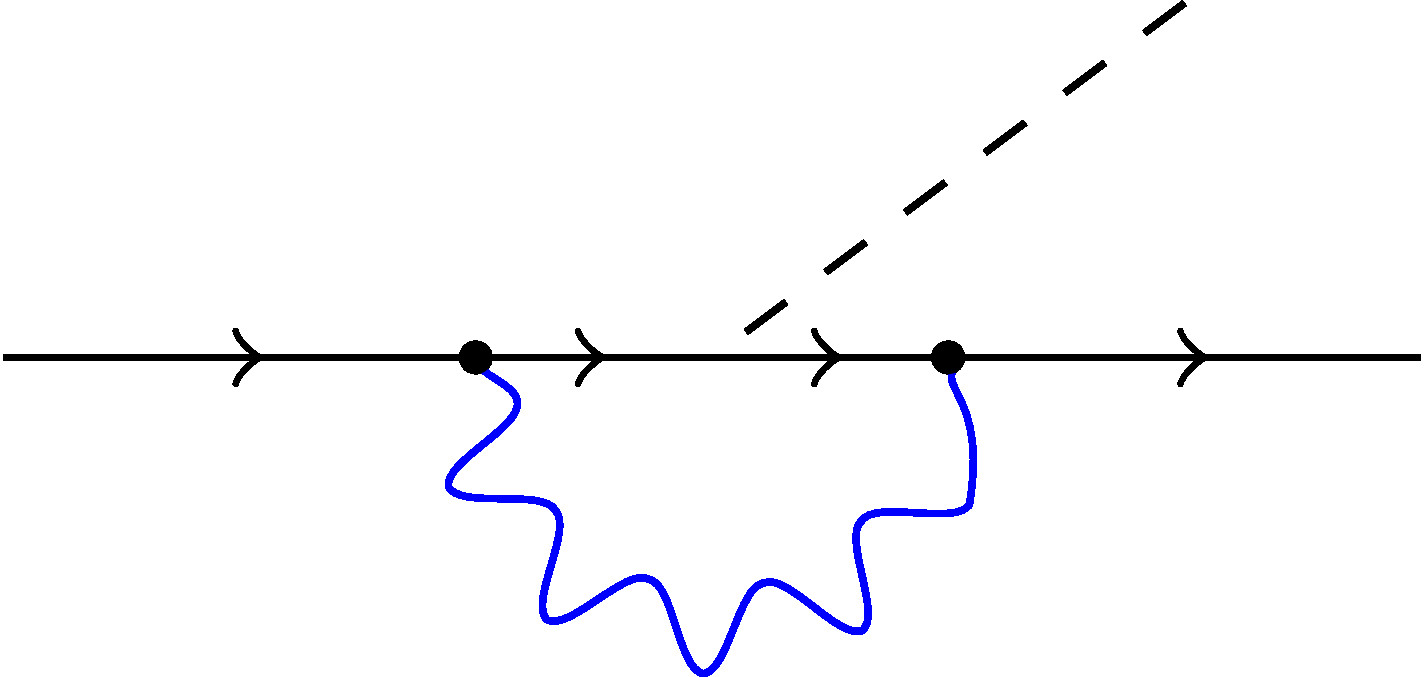}}\ + \ 
\raisebox{-0.6cm}{\includegraphics{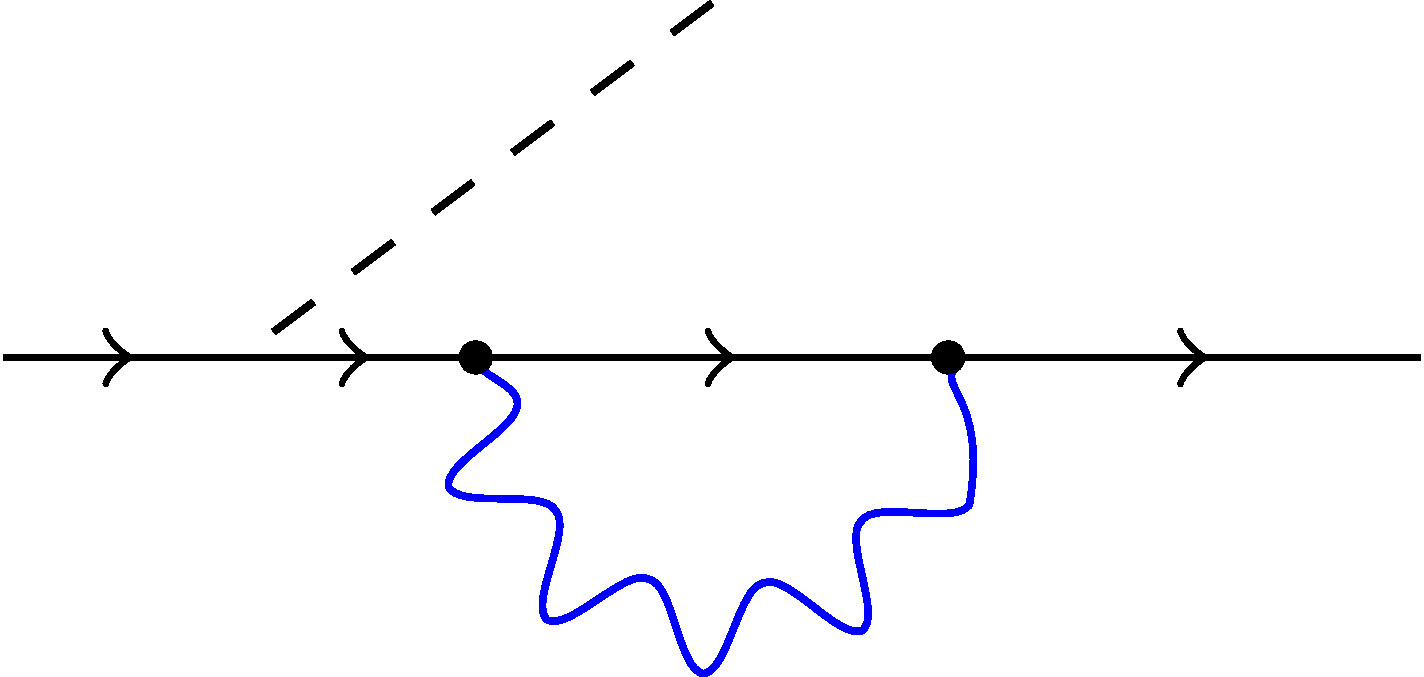}}
	\]
	\caption{Single emission with a loop correction.}
	\label{Fig:out}
\end{figure}

We now turn to the one loop corrections to the basic interactions between the matter and its background. For the absorption process in Fig.~\ref{Fig:Ab} we have, at one loop, the three diagrams in Fig.~\ref{Fig:In},
and, for the emission process of Fig.~\ref{Fig:Em}, we get the contributions in Fig.~\ref{Fig:out}.
Note that the central term for each row here has the structure of a vertex correction, while the other terms are self-energies for the external legs. So it is not immediately clear that grouping them together in this way leads to a multiplicative renormalisation of the tree level processes in Figs.~\ref{Fig:Ab} and~\ref{Fig:Em}.

To clarify how renormalisation works in this context, we first need to recall how  sideband structures emerge from the tree level diagrams in Figs.~\ref{Fig:Ab} and~\ref{Fig:Em}. To that end, we note that the absorption process of~Fig.~\ref{Fig:Ab} can be written as the difference of two  propagators:
\begin{equation}\label{eq:Sb1}
\Prop{n+1}\Ab\Prop{n}=\In\Prop{n}-\Prop{n+1}\In
\end{equation} 
where we define the \lq In\rq\ term as
\begin{equation}\label{eq:In}
  \In=\frac{2\plaser+\ksl\lasersl}{2\pk}\,.
\end{equation}
The simple identity~(\ref{eq:Sb1}) lies at the heart of the sideband description of this propagation that was introduced in~\cite{Reiss:1966A}. At its heart, it is simply a partial fraction expansion which relates the products of propagators to their sums. 
 
The corresponding emission  version of this sideband identity can be easily deduced by using the duality transformation (\ref{eq:dualitytree}) and (\ref{eq:Sb1}) to give
\begin{equation}\label{eq:Sb2}
\Prop{n}\Em\Prop{n+1}=-\overline{\Prop{n+1}\Ab\Prop{n}}=\Prop{n}\Out-\Out\Prop{n+1}\,, 
\end{equation}
where the \lq Out\rq\ insertion is given by
\begin{equation}\label{eq:Out}
   \Out:=\overline{\In}=\frac{2\plasers-\ksl\laserssl}{2\pk}\,.
\end{equation}
The matrix nature of the $\In$ and $\Out$ terms means that we must be careful with the ordering in (\ref{eq:Sb1}) and (\ref{eq:Sb2}). However, due to the null nature of $\ksl$ and the fact that it commutes with both $\lasersl$ and $\laserssl$, we find that the In and Out terms  commute: 
\begin{equation}\label{eq:OI_com}
   [\In,\Out]=0\,.
\end{equation} 
Having clarified the tree level structures in Figs.~\ref{Fig:Ab} and~\ref{Fig:Em}, we can now analyse in much the same way the loop corrections of Figs.~\ref{Fig:In} and~\ref{Fig:out}.

\begin{figure}[htb] 
	\[
\raisebox{-0.6cm}{\includegraphics{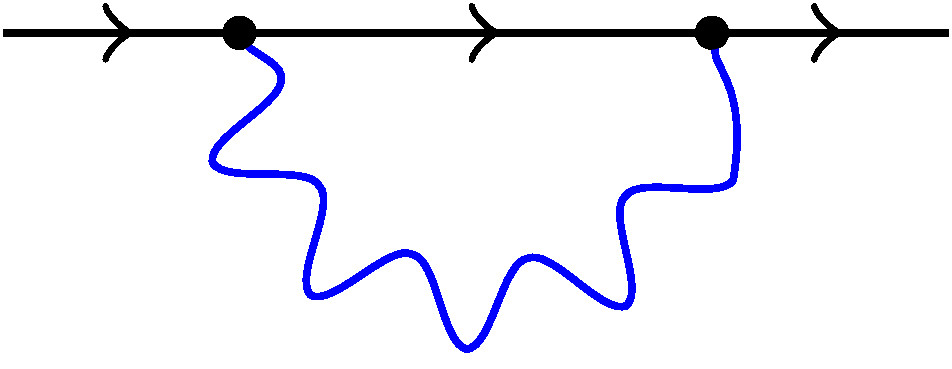}} 
=
\Prop{n}\Sigma_n\Prop{n}
	\]
	\caption{One loop self energy correction to the propagator. }
	\label{Fig:Z2}
\end{figure}

The ultraviolet poles related to the self energy contributions of Fig.~\ref{Fig:In} can  be readily calculated by using standard results from QED, see for example chapter~18 of~\cite{Schwartz:2013pla}. Working in the Feynman gauge, and using dimensional regularisation in $D=4-2\varepsilon$ dimensions, we have  for incoming momentum~$p+nk$ the contribution described in Fig.~\ref{Fig:Z2}.
After some simplifications of the gamma matrices, we have for the ultraviolet divergent structure
\begin{equation}
\Sigma_n=-e^2\mu^{2\ve}	\intfsDuv\frac{(2-D)\ssl+Dm}{(s-(p+nk))^2(s^2-m^2)}\,,
\end{equation}
where $s$ is the four-momentum of the electron in the loop so that the photon in the loop has four-momentum $p+nk-s$. 

Retaining only the ultraviolet pole  gives
\begin{equation}\label{eq:Sigma_n}
  \Sigma_n = (i3m+\PropI{n})\duv \,.
\end{equation}
The notation here is that, from (\ref{eq:Prop}), $\PropI{n}=(-i)(\psl+n\ksl-m)$ while the ultraviolet pole is given by
\begin{equation}\label{eq:deltaUV}
  \duv=-\frac{e^2}{(4\pi)^2}\frac1{\varepsilon}\,.
\end{equation}
Substituting the  self-energy expression~(\ref{eq:Sigma_n}) into Fig.~\ref{Fig:Z2} gives the familiar double pole mass term 
and a single pole. So the first and third diagrams in Fig.~\ref{Fig:In} become
\begin{align}
\Prop{n+1}\Sigma_{n+1}&\Prop{n+1}\Ab\Prop{n}+\Prop{n+1}\Ab\Prop{n}\Sigma_{n}\Prop{n}\\\nonumber
&=\In\Prop{n}\Sigma_{n}\Prop{n}-\Prop{n+1}\Sigma_{n+1}\Prop{n+1}\In+\Prop{n+1}\Sigma_{n+1}\In\Prop{n}-
\Prop{n+1}\In\Sigma_{n}\Prop{n}\,.
\end{align}
The first two terms on the right hand side here are the sideband structures but the final two include a mixture of momenta. 

 \begin{figure}[htb]
	\[
\raisebox{-0.6cm}{\includegraphics{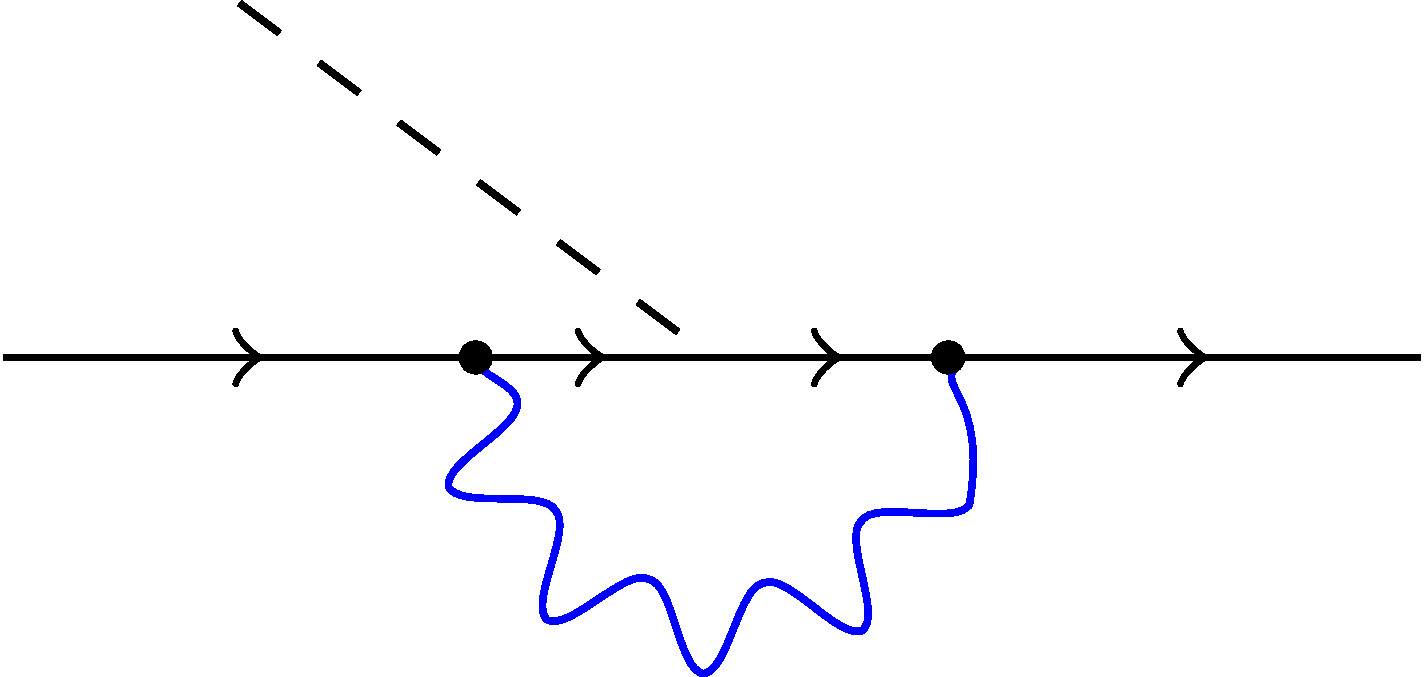}} 
=
\Prop{n+1} \Sigma_{\mathrm{in}}\Prop{n}
	\]
	\caption{The one loop absorption vertex correction to the propagator.}
	\label{Fig:Z2a}
\end{figure}

The vertex correction term in  Fig.~\ref{Fig:In} is still to be included.  The corresponding Feynman rule for this is given in Fig.~\ref{Fig:Z2a}. 
From this we have
\begin{equation}
\Sigma_{\mathrm{in}}=-e^2\mu^{2\ve}	\intfsDuv\frac{\gamma^\rho (\ssl+\ksl+m)\lasersl(\ssl+m)\gamma^\tau}{((s+k)^2-m^2)(s^2-m^2)}\frac{g_{\rho\tau}}{(s-(p+nk))^2}\,.
\end{equation}
Retaining only the ultraviolet divergent structures, which can easily be recognised by power counting, we find
\begin{equation}\label{eq:inraw}
  \Sigma_{\mathrm{in}}=  i \lasersl\,\duv=-\Ab\duv\,.
\end{equation}
Here we recognise the tree level absorption factor of Fig.~\ref{Fig:Ab} multiplied by the above ultraviolet pole. We emphasise that, in the last step, the $e^2$ factor from the loop is in the $\duv$ term, while the background coupling factor of $e$ has been absorbed into the definitions of~$\lasersl$ and~$\Ab$.

This simple relation for the ultraviolet structure of this vertex term means that we can exploit the sideband relation~(\ref{eq:Sb1}) to rewrite  
\begin{align}
  \Sigma_{\mathrm{in}}&=\In(\duv\Prop{n}^{-1})-(\duv\Prop{n+1}^{-1})\In\,,\\\nonumber
  &=\In\Sigma_n-\Sigma_{n+1}\In\,.
\end{align}
Note that the  scalar mass terms cancelled in the last step. 
The second diagram in Fig.~\ref{Fig:In} can thus be written as
\begin{equation}\label{eq:into2}
  \Prop{n+1}\Sigma_{\mathrm{in}}\Prop{n}=\Prop{n+1}\In \Sigma_n\Prop{n}-\Prop{n+1} \Sigma_{n+1}\In\Prop{n}\,.
\end{equation}

We can now write the sum of the  three diagrams in Fig.~\ref{Fig:In} as
\begin{align}
\begin{split}
  \Prop{n+1}\Sigma_{n+1}&\Prop{n+1}\Ab\Prop{n}+\Prop{n+1}\Sigma_{\mathrm{in}}\Prop{n} + \Prop{n+1}\Ab\Prop{n}\Sigma_{n}\Prop{n}\\
  &=\In\Prop{n}\Sigma_n\Prop{n}-\Prop{n+1}\Sigma_{n+1}\Prop{n+1}\In\,.
\end{split}
\end{align} 
Note that all of the non-sideband structures cancel. What remains has exactly the same structure as the sideband description of the tree level result (\ref{eq:Sb1}), but with the expected self-energy corrections to the sideband propagators. 
We thus see the attractive result that the loop corrections to (\ref{eq:Sb1}) generate  the normal one-loop propagator  corrections to the tree level propagators in the sidebands:
\begin{equation}\label{eq:ren_in}
  \In\Big(\Prop{n}+\Prop{n}\Sigma_{n}\Prop{n}\Big)-
  \Big(\Prop{n+1}+\Prop{n+1}\Sigma_{n+1}\Prop{n+1}\Big)\In\,.
\end{equation}
The interpretation of this result is then direct: it will lead to the sidebands  requiring the standard  mass and wave function renormalisations.  

The emission process of Fig.~\ref{Fig:Em}, and its loop corrections in Fig.~\ref{Fig:out}, then lead to the sidebands described in (\ref{eq:Sb2}) being renormalised in a similar way. The key out-going vertex identity, dual  to (\ref{eq:into2}), is that
\begin{equation}\label{eq:outto2}
  \Prop{n}\Sigma_{\mathrm{out}}\Prop{n+1}=\Prop{n}\Sigma_n\Out\Prop{n+1}-\Prop{n}\Out \Sigma_{n+1}\Prop{n+1}\,,
\end{equation}
where we have used $\overbar{\Sigma}_{\mathrm{in}}=-\Sigma_{\mathrm{out}}$ and $\overbar{\Sigma}_{n}=-\Sigma_{n}$.
This then results in the loop corrections to the sidebands (\ref{eq:Sb2}) being given by
\begin{equation}\label{eq:ren_out}
  \Big(\Prop{n}+\Prop{n}\Sigma_{n}\Prop{n}\Big)\Out-\Out 
  \Big(\Prop{n+1}+\Prop{n+1}\Sigma_{n+1}\Prop{n+1}\Big)\,.
\end{equation} 
This we see is precisely (minus) the dual of (\ref{eq:ren_in}) as we would naively expect from the tree level relation (\ref{eq:dualitytree}). Again, the standard mass and wave-function renormalisations will suffice.

\section{Higher order background corrections}
Having understood the structure of the loop correction to  a single absorption or emission of a laser photon by the electron, we now want to calculate the ultraviolet divergences when multiple photons are absorbed or emitted. We shall consider the case of both absorption and emission in the following section.

\begin{figure}[htb]
	\[
\raisebox{-0.6cm}{\includegraphics{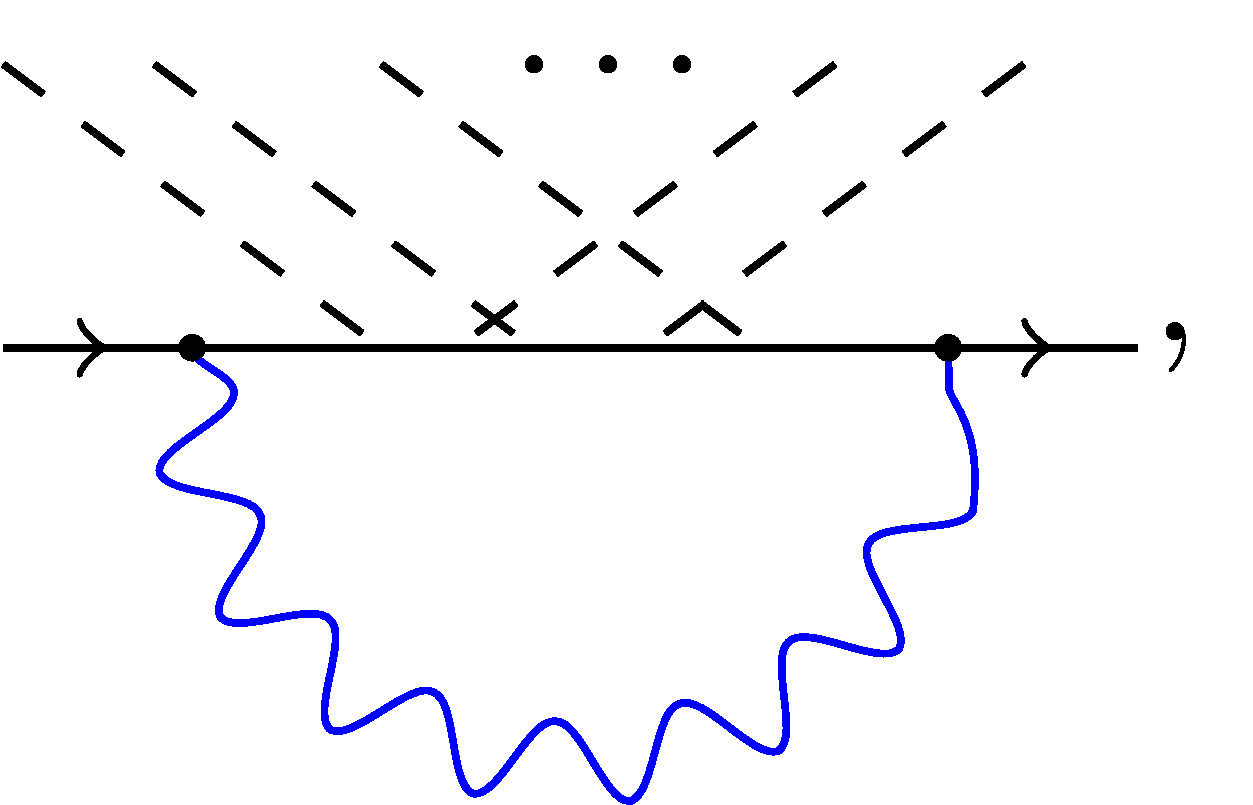}}
	\]
	\caption{Loop spanning multiple laser interactions.}
	\label{Fig:many_in_loop}
\end{figure}
The first thing to note is that loops spanning more than one laser absorption or emission, as depicted in Fig.~\ref{Fig:many_in_loop}, 
are all finite in the ultraviolet regime  by simple power counting.

\begin{figure}[htb]
	\[
\raisebox{0cm}{\includegraphics{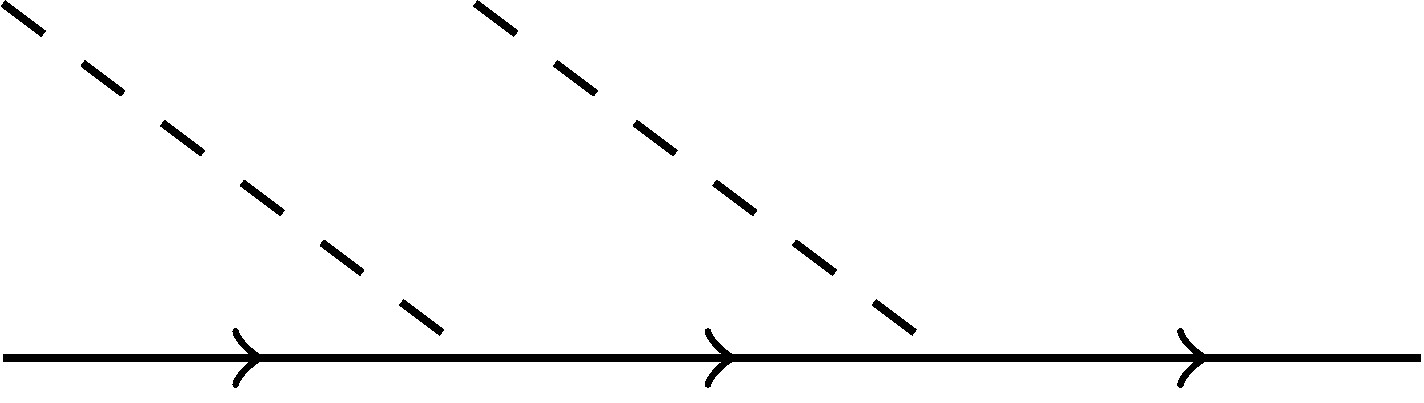}} =\Prop{n+2}\Ab\Prop{n+1}\Ab\Prop{n}
	\]
	\caption{Tree level double absorption process.}
	\label{Fig:inandin}
\end{figure}

\begin{figure}[htb]
	\[
\raisebox{-1cm}{\includegraphics{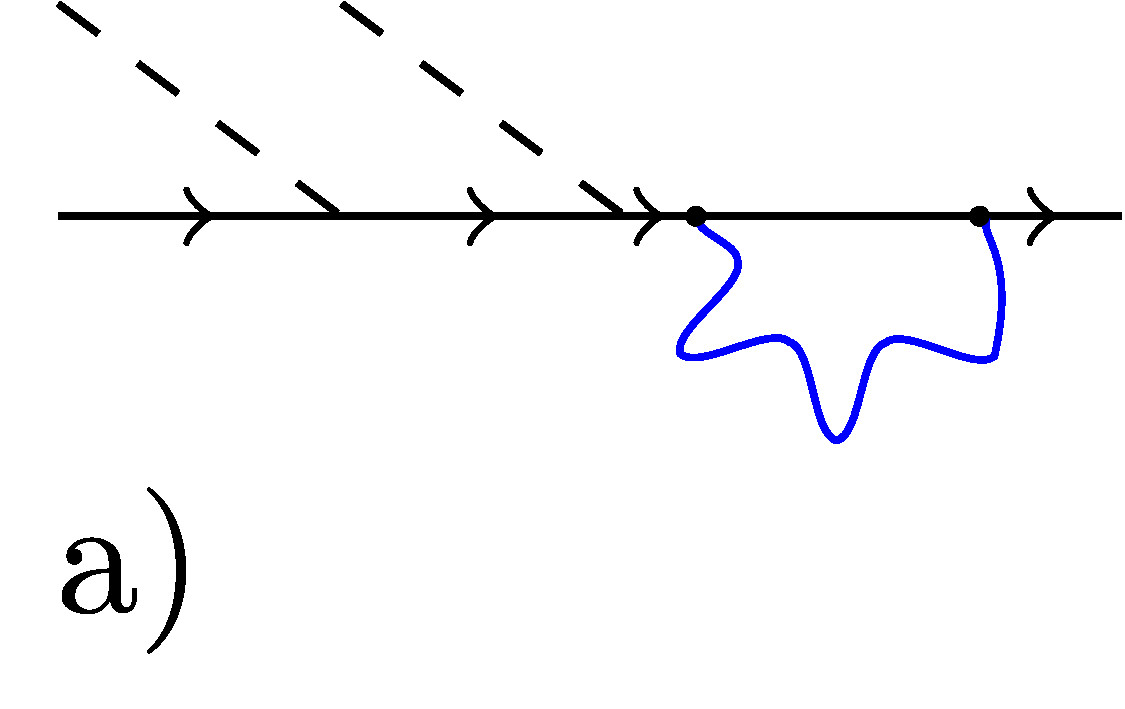}}\quad
\raisebox{-1cm}{\includegraphics{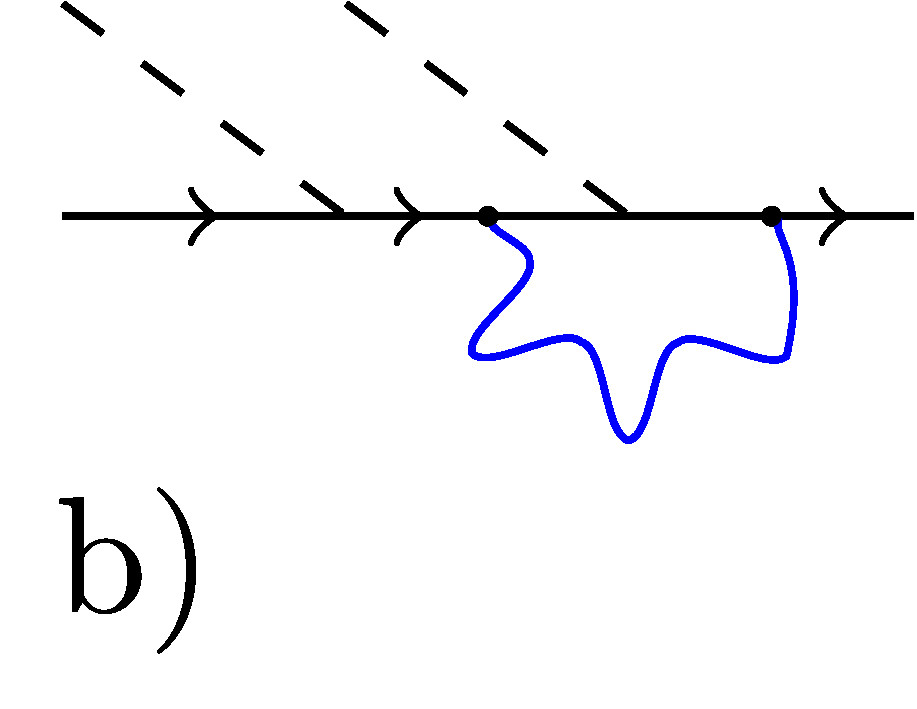}}\quad
\raisebox{-1cm}{\includegraphics{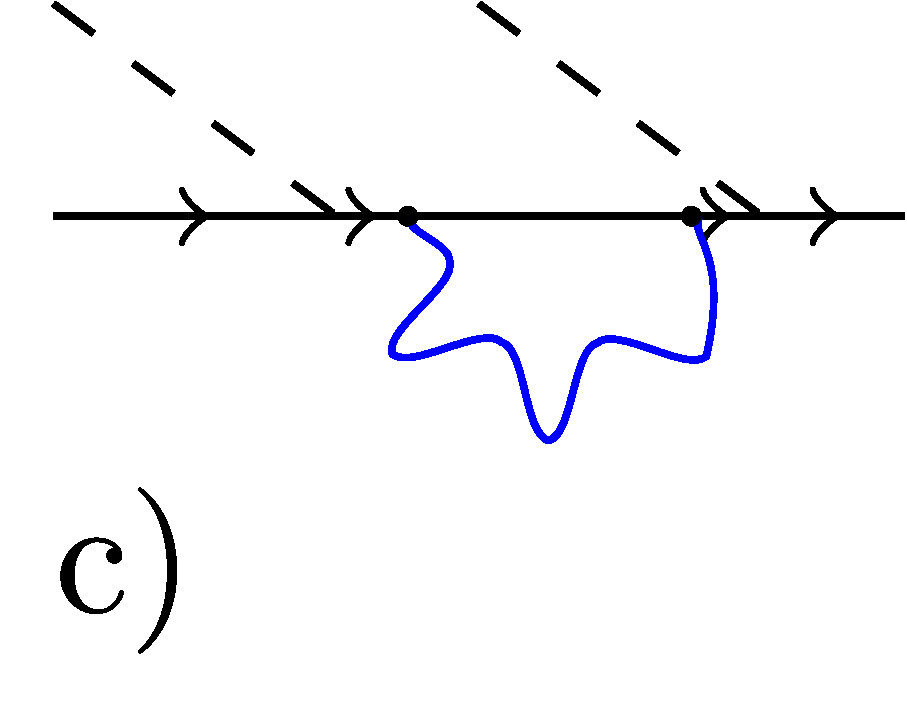}}\quad
\raisebox{-1cm}{\includegraphics{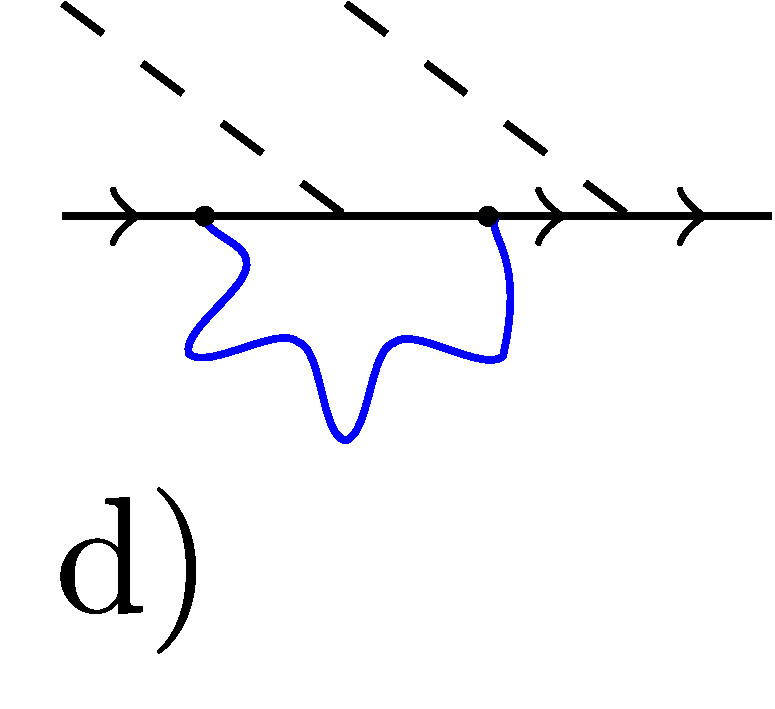}}\quad
\raisebox{-1cm}{\includegraphics{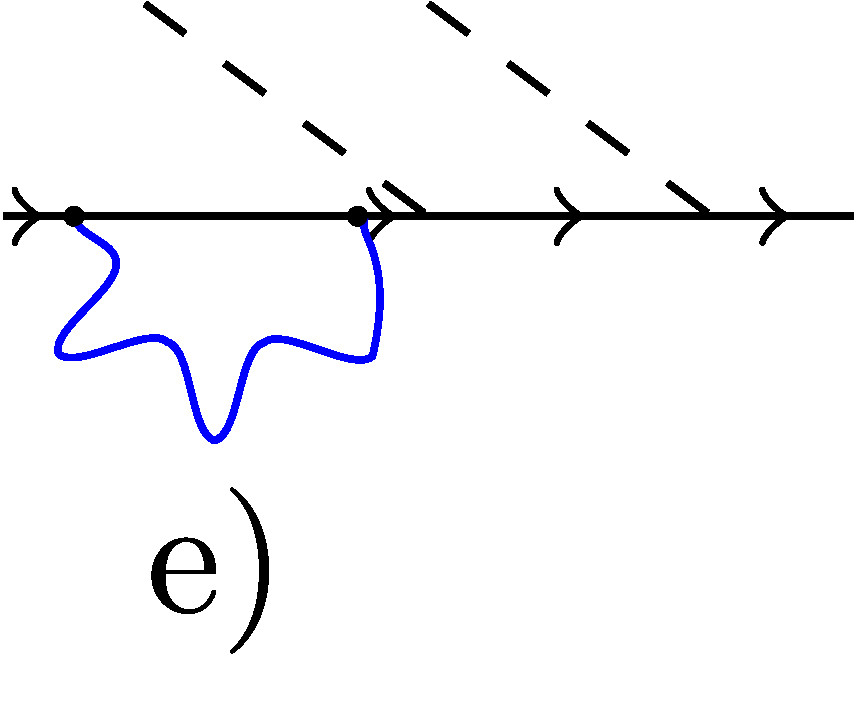}}
	\]
	\caption{Double absorption process with a loop correction.}
	\label{Fig:ininloop}
\end{figure}

This means that when, for example,  we consider the tree level double absorption process, where the incoming  propagator $\Prop{n}$ absorbs two additional laser photons, as in Fig.~\ref{Fig:inandin}, then we need only to consider the loop corrections straddling no more than one background vertex, as shown in Fig.~\ref{Fig:ininloop}.
In this we again see a mixture of self-energy and single vertex corrections. 

In order to understand and interpret these corrections, we need to first identify the sideband structures in the tree level term shown in Fig.~\ref{Fig:inandin}.  To that end, we write this as
\begin{equation}
  \Prop{n+2}\Ab\Prop{n+1}\Ab\Prop{n}=\Prop{n+2}\Ab\Prop{n+1}\PropI{n+1}\Prop{n+1}\Ab\Prop{n}\,.
\end{equation}
This allows us to use the absorption relation (\ref{eq:Sb1}) twice, resulting in four terms:
\begin{equation}\label{eq:2instep}
\Prop{n+2}\Ab\Prop{n+1}\Ab\Prop{n}=\Prop{n+2}\In^2-\In\Prop{n+1}\In+\In^2\Prop{n}-\Prop{n+2}\In\PropI{n+1}\In\Prop{n}\,.
\end{equation}
A key identity needed here, which is straightforward to show,  is that
\begin{equation}\label{eq:IPI}
  \In\PropI{n+1}\In=\big(\tfrac12\In^2+\tfrac12v\big)\PropI{n}+\PropI{n+2}\big(\tfrac12\In^2-\tfrac12v\big)
\end{equation}
where 
\begin{equation}\label{eq:vdef}
  v:=\frac{\laserv}{2\pk}\,.
\end{equation}
Using this identity in (\ref{eq:2instep}), we see that the sidebands for the double absorption process depicted in Fig.~\ref{Fig:inandin} are given by
\begin{equation}\label{eq:ininsb}
 \Prop{n+2}\Ab\Prop{n+1}\Ab\Prop{n}=\big(\tfrac12\In^2-\tfrac12v\big)\Prop{n}-\In\Prop{n+1}\In+\Prop{n+2}\big(\tfrac12\In^2+\tfrac12v\big)\,. 
\end{equation}
We will now show that the one loop diagrams in Fig.~\ref{Fig:ininloop} generate the expected, ultraviolet  one-loop corrections to these three sidebands. 

The loop correction  can  be evaluated by recognising in the diagrams of Fig.~\ref{Fig:ininloop}  a connection to the earlier loop processes evaluated in the previous section. The first three diagrams represent an initial absorption process followed by the loop corrections of Fig.~\ref{Fig:In}, with shifted initial momentum. In a similar way, the final three diagrams in Fig.~\ref{Fig:ininloop} can be interpreted as the  loop corrections of Fig~\ref{Fig:In}, followed immediately by an absorption process. These two simplifications  double count the middle process, Fig.~\ref{Fig:ininloop}c, so this needs to be subtracted from the combined sum. 

Following this reduction prescription, the diagrams in Fig.~\ref{Fig:ininloop}  can then be evaluated using the loop results (\ref{eq:ren_in}) and the sideband  identity (\ref{eq:Sb1}). This results in terms containing combinations of the form $\In\Sigma_{n}\In$ which, from the self-energy extension to (\ref{eq:IPI}), can be evaluated by using the  identity 
 \begin{equation}\label{eq:ISigmaI}
  \In\Sigma_{n+1}\In=\big(\tfrac12\In^2+\tfrac12v\big)\Sigma_{n}+\Sigma_{n+2}\big(\tfrac12\In^2-\tfrac12v\big)\,.
\end{equation}
From this we rapidly arrive at the sideband structure of the one loop corrections of Fig.~\ref{Fig:ininloop}  to the  double absorption process shown in Fig.~\ref{Fig:inandin}. Combined with the tree-level result, this yields
\begin{align}\label{eq:ren_inin}
\begin{split}
  \big(\tfrac12\In^2-\tfrac12 v\big)&\Big(\Prop{n}+\Prop{n}\Sigma_{n}\Prop{n} \Big)-\In\Big(\Prop{n+1}+\Prop{n+1}\Sigma_{n+1}\Prop{n+1}\Big)\In\\&+\Big(\Prop{n+2}+\Prop{n+2}\Sigma_{n+2}\Prop{n+2} \Big)\big(\tfrac12\In^2+\tfrac12 v\big)\,.
 \end{split}
\end{align}
Again we see that the sidebands pick up the expected  loop corrections.

\begin{figure}[htb]
	\[
\raisebox{-0.2cm}{\includegraphics{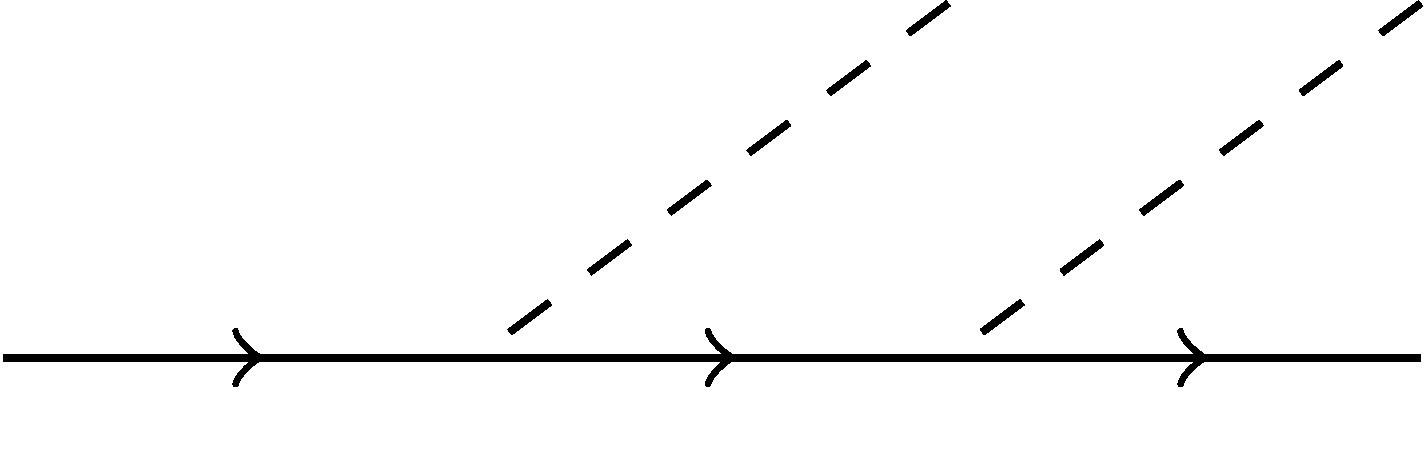}}=\Prop{n-2}\Em\Prop{n-1}\Em\Prop{n}
	\]
	\caption{Tree level double emission process.}
	\label{Fig:outandout}
\end{figure}

The double emission process can be evaluated in a similar fashion, or more directly by taking the dual of the double absorption process. The result is that the loop corrections to the double emission processes described in Fig.~\ref{Fig:outandout}
are given by the sideband terms:
\begin{align}\label{eq:ren_outout}
\begin{split}
  \Big(\Prop{n}+\Prop{n}\Sigma_{n}\Prop{n} \Big)&\big(\tfrac12\Out^2-\tfrac12 v^*\big)-\Out\Big(\Prop{n+1}+\Prop{n+1}\Sigma_{n+1}\Prop{n+1}\Big)\Out\\&+\big(\tfrac12\Out^2+\tfrac12 v^*\big)\Big(\Prop{n+2}+\Prop{n+2}\Sigma_{n+2}\Prop{n+2} \Big)\,,
 \end{split}
\end{align}
where now
\begin{equation}\label{eq:vsdef}
  v^*:=\frac{\laservs}{2\pk}\,.
\end{equation}
An important point to note here is that the terms $v$ and $v^*$ induced by the background do not acquire  loop corrections and hence are not renormalised at one loop. We also  note that both $v$ and $v^*$ are polarisation dependent and vanish for a circularly polarised laser, see~\cite{Lavelle:2017dzx}.

\section{Absorption and emission from the background} 
It is well known that the laser induced mass shift is only generated by  processes where there is both emission and absorption from the laser. This is understood at all orders in the background field and is known to be polarisation independent, see~\cite{Lavelle:2017dzx} and references therein. Here we will calculate the one-loop corrections to this important process and see the necessity for a new renormalisation.

\begin{figure}[htb]
	\[
\raisebox{-0.1cm}{\includegraphics{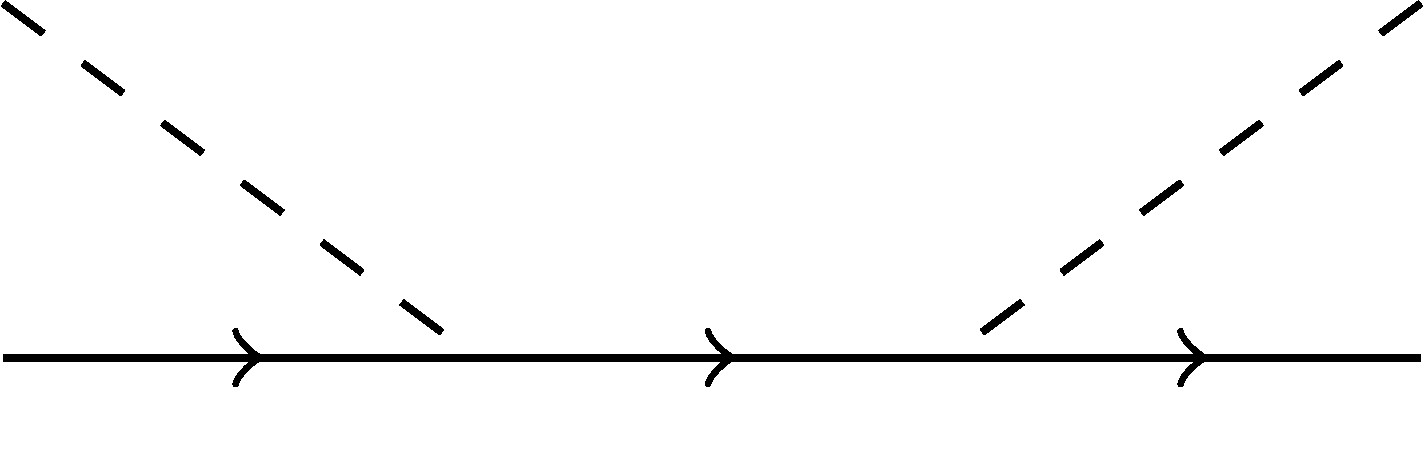}}
+
\raisebox{-0.1cm}{\includegraphics{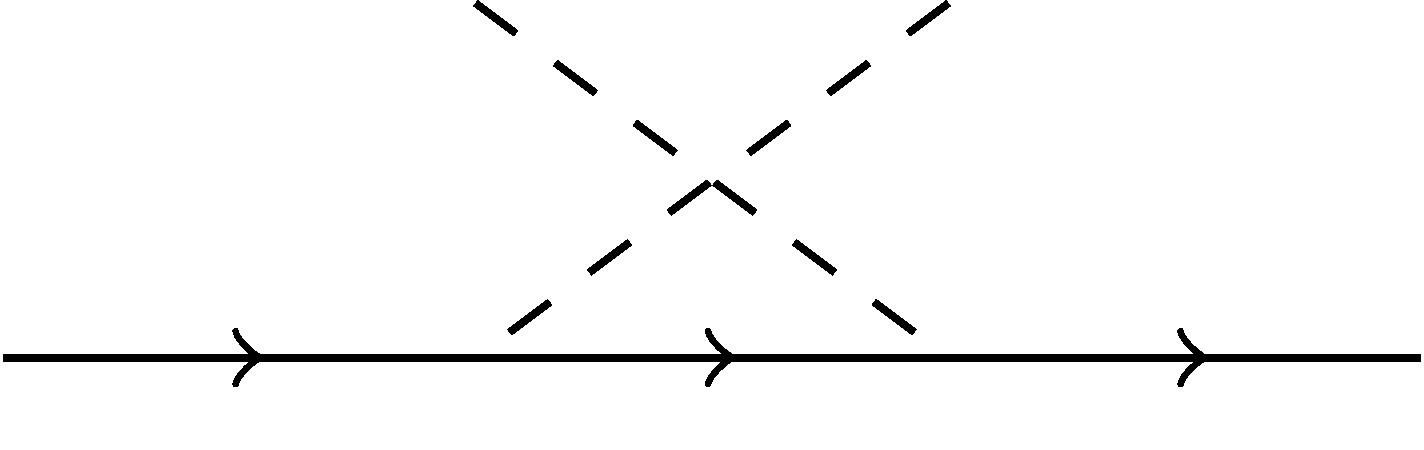}}
	\]
	\caption{Tree level absorption  and emission corrections.}
	\label{Fig:inout}
\end{figure}

There are two contributions to the mixed absorption and emission process at the lowest order in the background interactions, as summarised in Fig.~\ref{Fig:inout}.
We expect from~\cite{Lavelle:2015jxa} that these diagrams will generate three sidebands and the central one will involve a double pole corresponding to the laser induced mass shift. 

If the incoming momentum is again $p+nk$, then these diagrams are given by
\begin{equation}\label{eq:inout}
  \Prop{n}\Em\Prop{n+1}\Ab\Prop{n}+\Prop{n}\Ab\Prop{n-1}\Em\Prop{n}\,.
\end{equation}
Note that each of these contributions is unchanged (up to a sign) by the duality transformations introduced earlier.

The terms in (\ref{eq:inout}) can be evaluated by again inserting appropriate inverse propagators so that both the absorption and emission identities, (\ref{eq:Sb1}) and (\ref{eq:Sb2}), can be used. From this we quickly find that
\begin{align}
   \Prop{n}\Em\Prop{n+1}\Ab\Prop{n}+\Prop{n}\Ab\Prop{n-1}\Em\Prop{n}&=\In\Prop{n-1}\Out-2\Out\In\Prop{n}-\Prop{n}2\Out\In+\Out\Prop{n+1}\In\\\nonumber
   &\qquad +\Prop{n}(\Out\PropI{n+1}\In+\In\PropI{n-1}\Out)\Prop{n}\,.
\end{align}
The first four terms in this involve the expected sidebands for these processes, but the coefficients are not as expected. 
The final term needs more work to be interpreted, but should correct these coefficients. 

The structure in the brackets in the last equation is  analogous to the double absorption contribution seen earlier in (\ref{eq:IPI}). The key identity now is that 
\begin{equation}\label{eq:massid}
  \Out\PropI{n+1}\In+\In\PropI{n-1}\Out=\Out\In\PropI{n}+\PropI{n}\Out\In-i\mstarsl\,,
\end{equation} 
where we define the important quantity
\begin{equation}\label{eq:vector_mass}
  \mstar_\mu:
  =-\frac{\laserm}{\pk}k_\mu\,. 
\end{equation}
Note that $\mstarsl=\overbar{\mstarsl}$.
 
Using (\ref{eq:massid}) we find that the sidebands for this process are given at this order by:
\begin{equation}\label{eq:central_sidebands}
\Prop{n}\Em\Prop{n+1}\Ab\Prop{n}+\Prop{n}\Ab\Prop{n-1}\Em\Prop{n}= \In\Prop{n-1}\Out-\Out\In\Prop{n}-\Prop{n}\Out\In-\Prop{n} i\mstarsl\Prop{n}+\Out\Prop{n+1}\In\,. 
\end{equation}
Here we recognise the expected three sidebands, $\Prop{n}$, $\Prop{n\pm1 }$, and the double pole. These terms must be interpreted as corrections, induced by the laser, to the  free propagator in the central sideband, $\Prop{n}$.

Since  $\mstarsl$ is in the double pole term for the central sideband, we can relate it to the more familiar polarisation independent effective mass, $m_{*}$, induced by the  background. Following the discussion in~\cite{Lavelle:2017dzx}, we can write at this order in the laser background, $\Prop{n}-\Prop{n}i\mstarsl\Prop{n}$ as 
\begin{align}\label{eq:propmM}
\begin{split}
  \frac{i}{\psl+n\ksl-m+i\epsilon}&+\frac{1}{\psl+n\ksl-m+i\epsilon}\mstarsl\frac{i}{\psl+n\ksl-m+i\epsilon}\,,\\
  &\approx\frac{i}{\psl+n\ksl-(m+\mstarsl)+i\epsilon}\,,\\
  &=\frac{i(\psl+n\ksl+m-\mstarsl)}{(p+nk)^2-m^2_{*}+i\epsilon}\,,
\end{split}
\end{align}
where  
\begin{equation}\label{eq:mstar}
  m^2_{*}=m^2+\psl\mstarsl+\mstarsl\psl=m^2-2 \laserm\,.
\end{equation}
Note that the last result is often rewritten as $m^2_{*}=m^2-e^2a^2$ where $-a^2>0$ is the amplitude squared of the background.  

\begin{figure}[htb]
	\[
\includegraphics{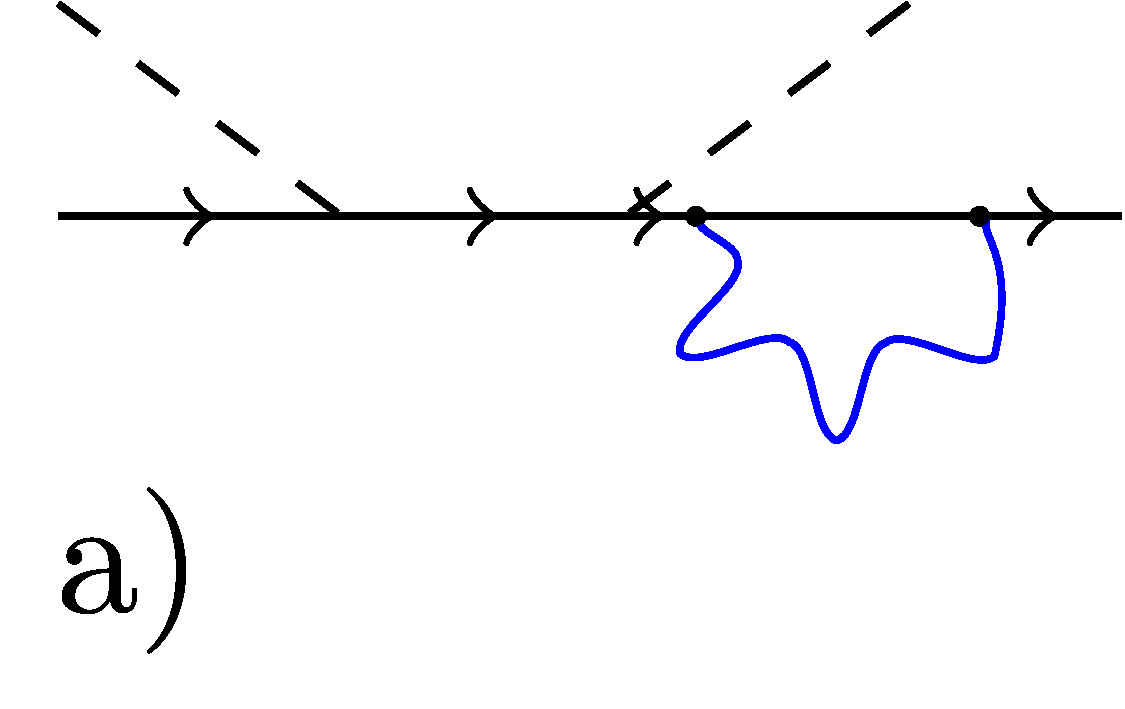}\quad
\includegraphics{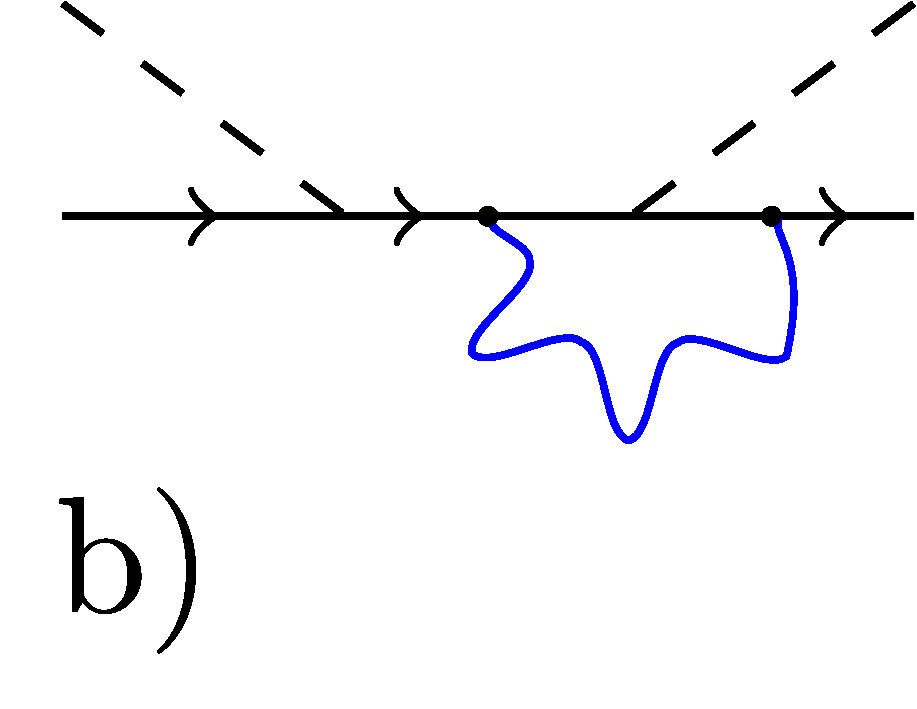}\quad
\includegraphics{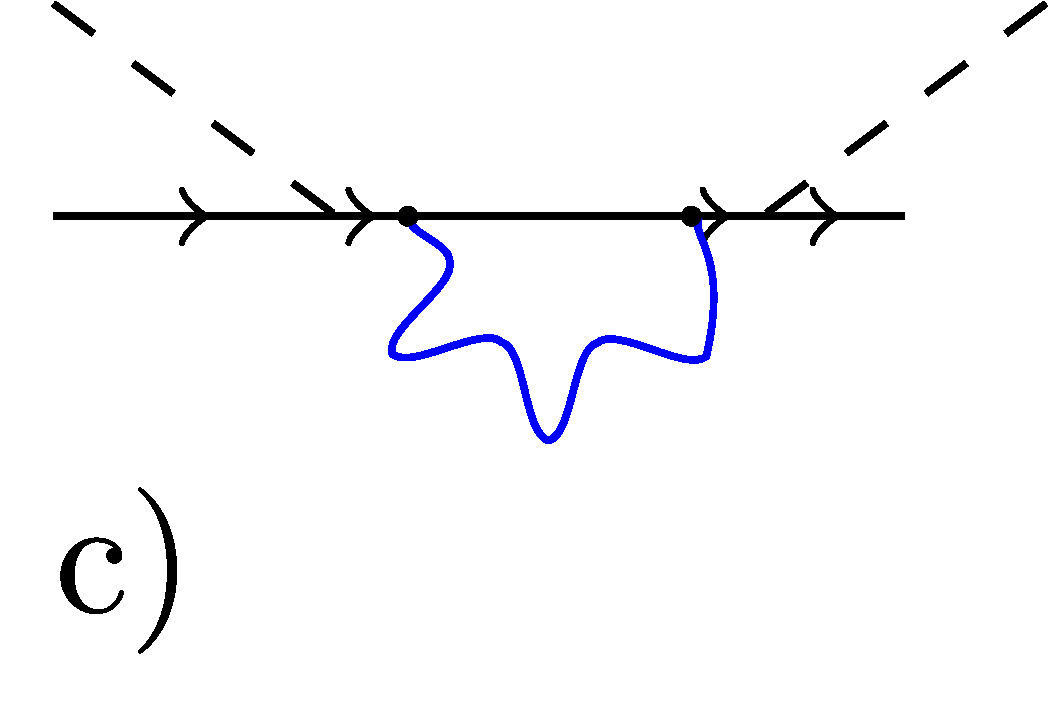}\quad
\includegraphics{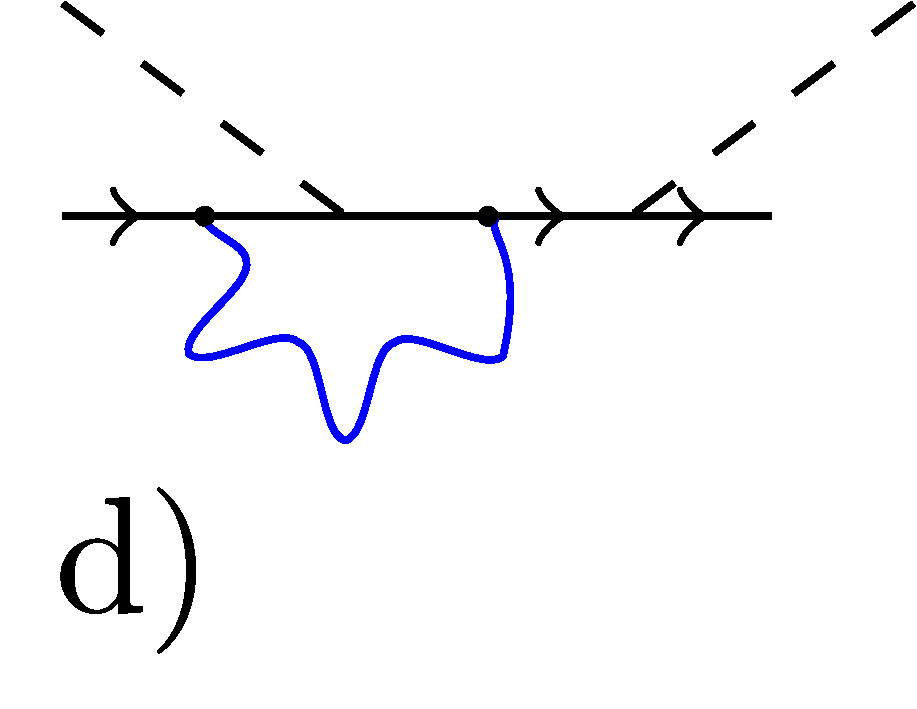}\quad
\includegraphics{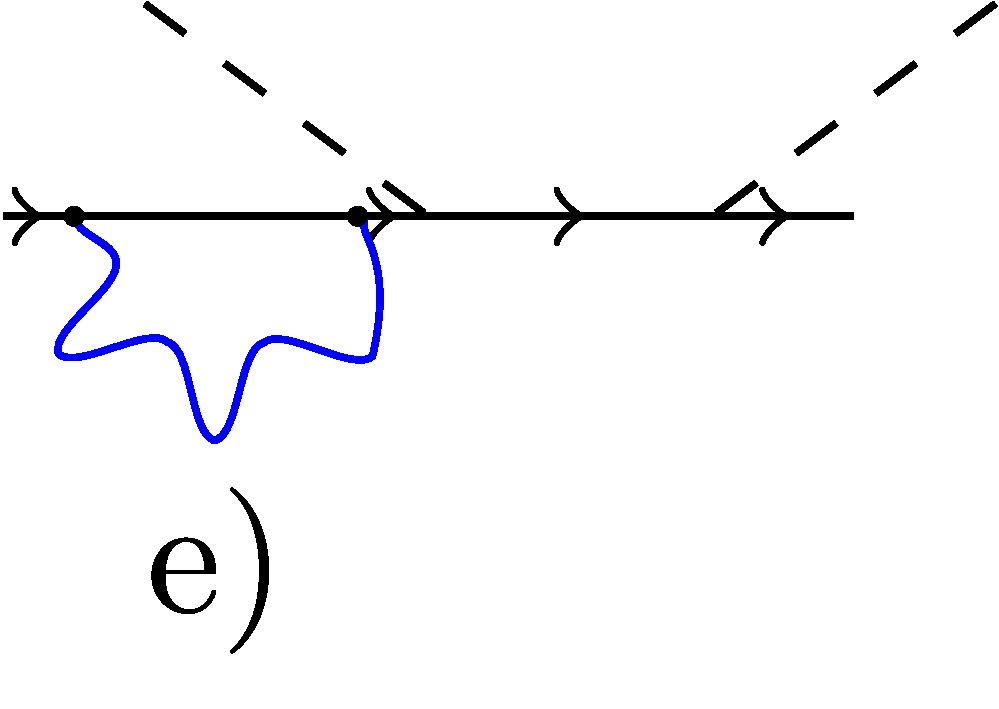}
	\]
	\caption{Absorption then emission with a loop correction.}
	\label{Fig:inoutloop}
\end{figure}

\begin{figure}[htb]
	\[
\includegraphics{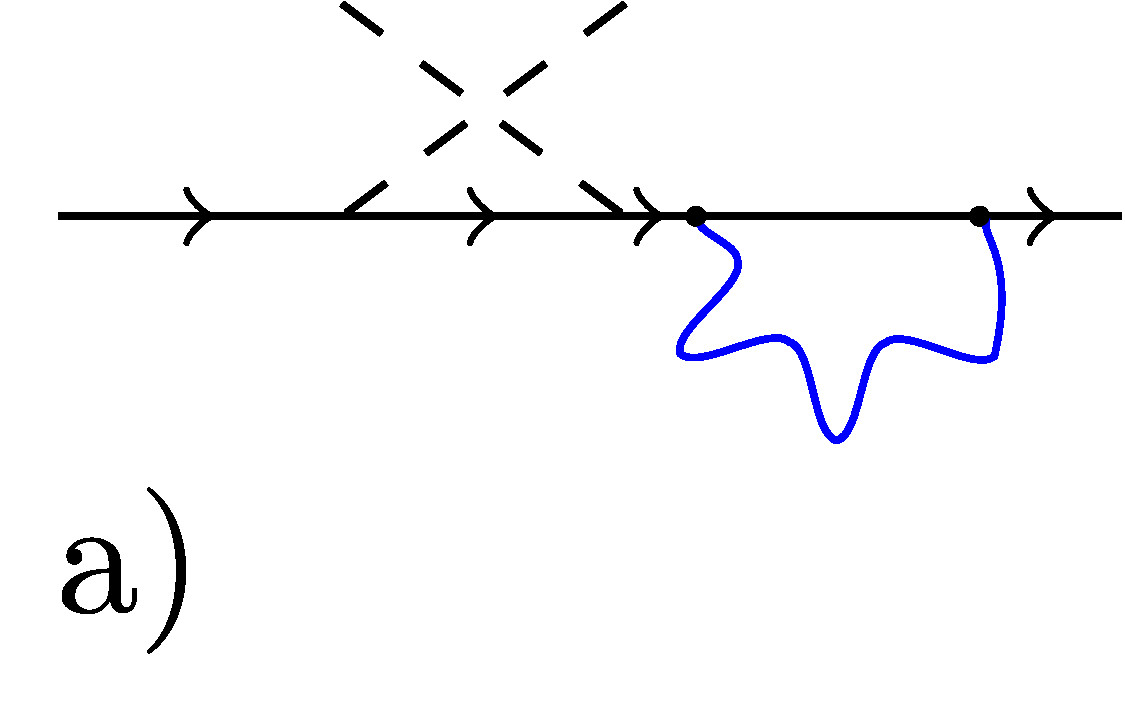}\quad
\includegraphics{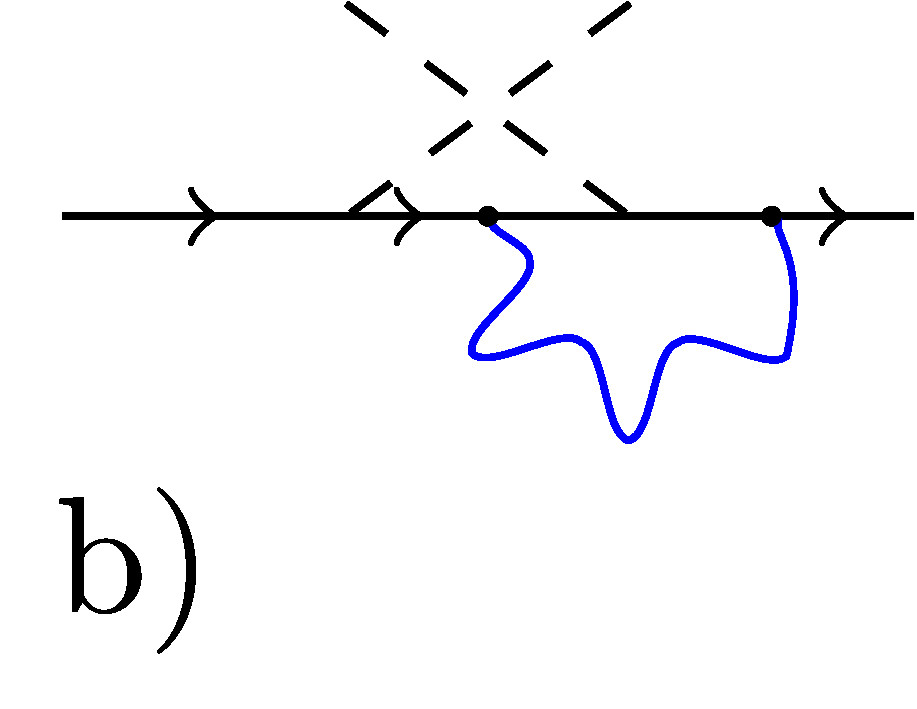}\quad
\includegraphics{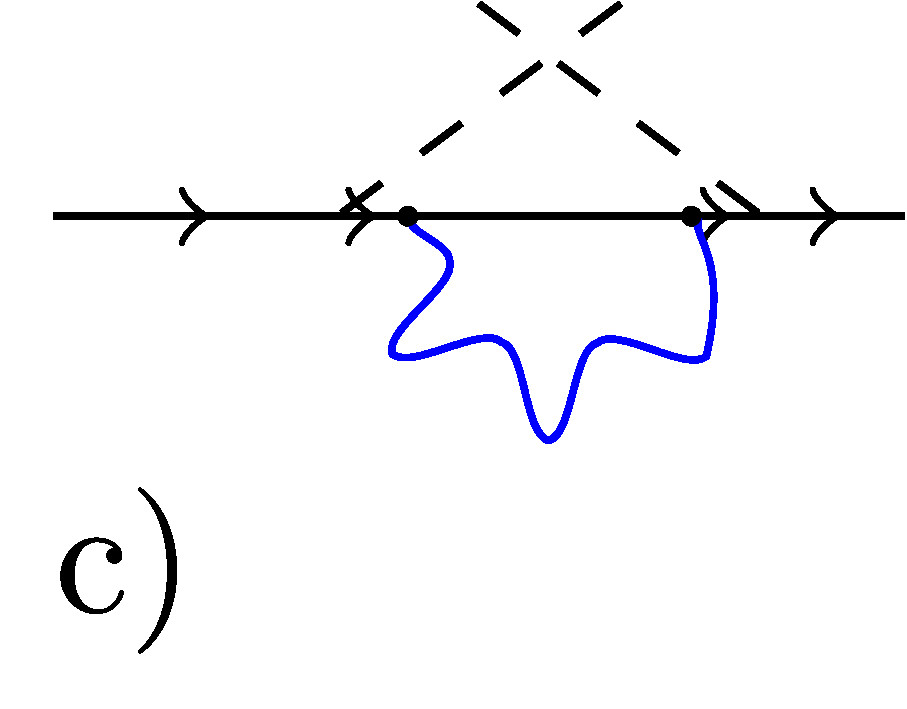}\quad
\includegraphics{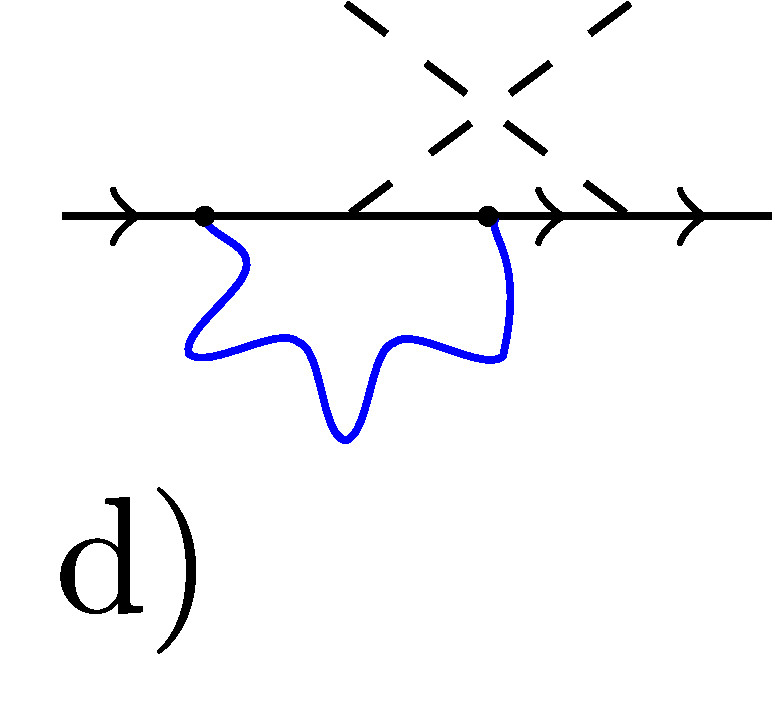}\quad
\includegraphics{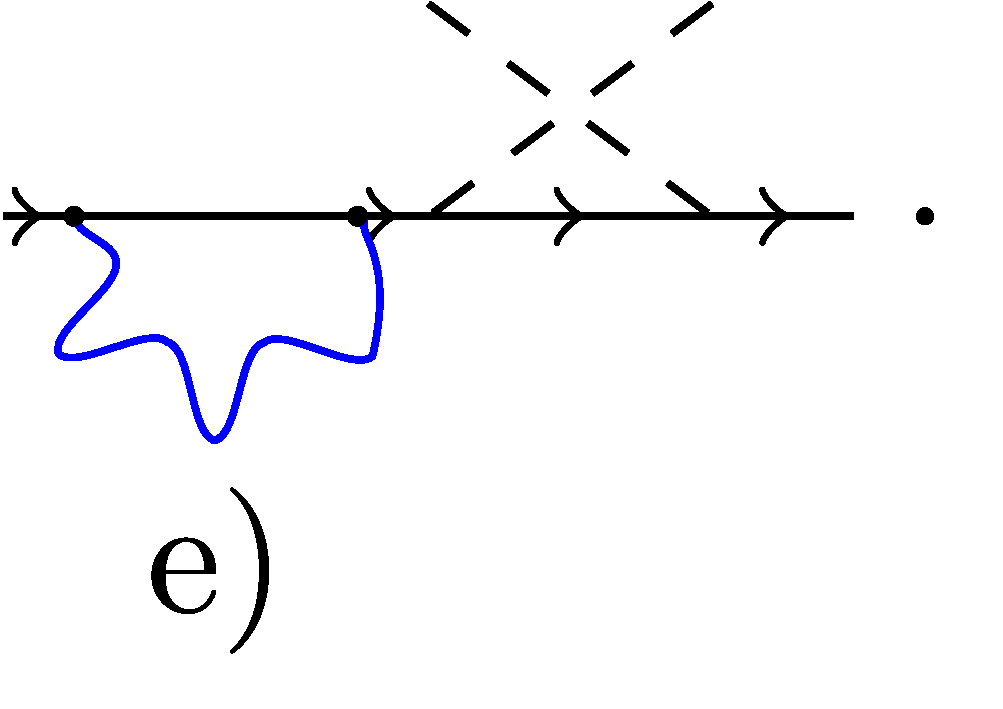}
	\]
	\caption{Emission then absorption with a loop correction.}
	\label{Fig:outinloop}
\end{figure}

We now want to calculate the  ultraviolet loop corrections to these processes. They  are given by the five diagrams in Fig.~\ref{Fig:inoutloop}
and the corresponding terms in Fig.~\ref{Fig:outinloop}.
Again we stress that since we are only calculating the ultraviolet divergences, loops straddling two laser lines may be ignored.

The strategy for evaluating these diagrams mirrors that seen before: we can identify sub terms that have already been evaluated,  then use the loop generalisation of the identity  (\ref{eq:massid}), which is 
\begin{equation}
  \Out\Sigma_{n+1}\In+\In\Sigma_{n-1}\Out=\Out\In\Sigma_{n}+\Sigma_{n}\Out\In+i\sigmasl\,,
\end{equation}
where we have defined the loop correction to the final term in (\ref{eq:massid}) 
\begin{equation}\label{eq:massUV}
  \sigmasl=\frac{e^2}{(4\pi)^2}\frac1\varepsilon \mstarsl
  \,.
\end{equation}
From this we find that the loop corrections to the central sidebands (\ref{eq:central_sidebands}) are given by (ignoring higher order terms in the coupling)
\begin{align}\label{eq:ren_inout}
\begin{split}
 &\In\Big(\Prop{n-1}+\Prop{n-1}\Sigma_{n-1}\Prop{n-1} \Big)\Out\\
 &\qquad-\Out\In\Big(\Prop{n}+\Prop{n}\Sigma_{n}\Prop{n} \Big) -\Big(\Prop{n}+\Prop{n}\Sigma_{n}\Prop{n} \Big)\Out\In\\
 &\qquad\qquad\qquad-\Big(\Prop{n}+\Prop{n}\Sigma_{n}\Prop{n}\Big)i\big(\mstarsl+\sigmasl\big)\Big(\Prop{n}+\Prop{n}\Sigma_{n}\Prop{n}\Big)\\
&\qquad\qquad+\Out\Big(\Prop{n+1}+\Prop{n+1}\Sigma_{n+1}\Prop{n+1} \Big)\In\,. 
\end{split}
\end{align} 
We have written it in this way to bring out the multiplicative structure of these corrections at this order. Terms without a $\Sigma$ are tree level, terms with one $\Sigma$ are our one loop results, while terms with products of two or more $\Sigma$ factors remain to be verified in further work as the calculations reported here are only to one loop. 

Written in this way, we see a new structure in the third line of the loop corrections: $\Prop{n}(-i\sigmasl)\Prop{n}$. This, as we will discuss in more detail later, corresponds to a new renormalisation being needed in this theory. This will be a renormalisation of the laser induced mass shift, $\mstarsl$. 

This last result is unexpected and requires further testing. To do this we now consider a process that is higher order in the background interaction and that also generates a laser induced mass shift at tree level. To be concrete, we will consider two absorptions and one emission.  

\begin{figure}[htb]
	\[
\raisebox{-.1cm}{\includegraphics{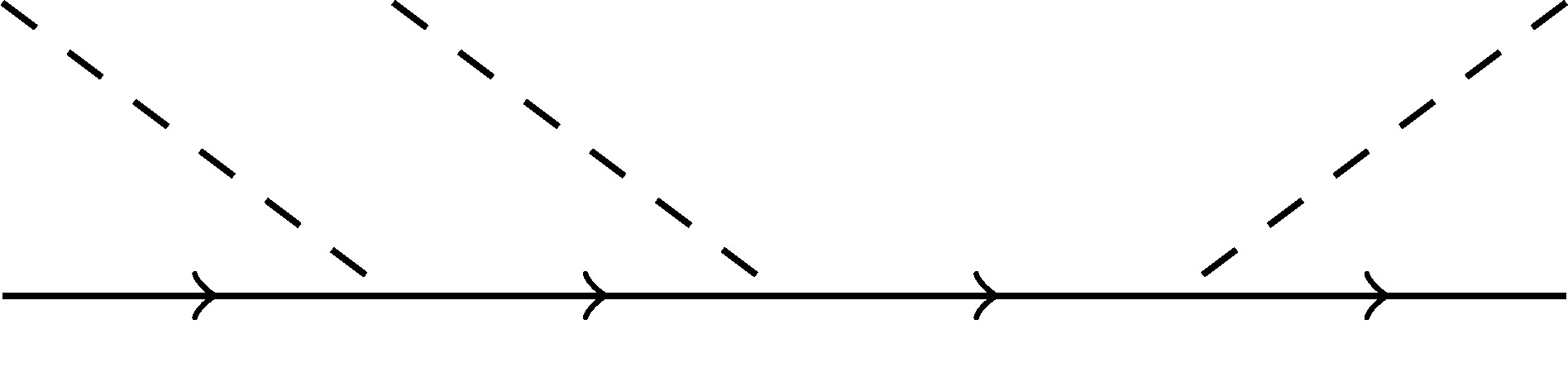}}
+
\raisebox{-.1cm}{\includegraphics{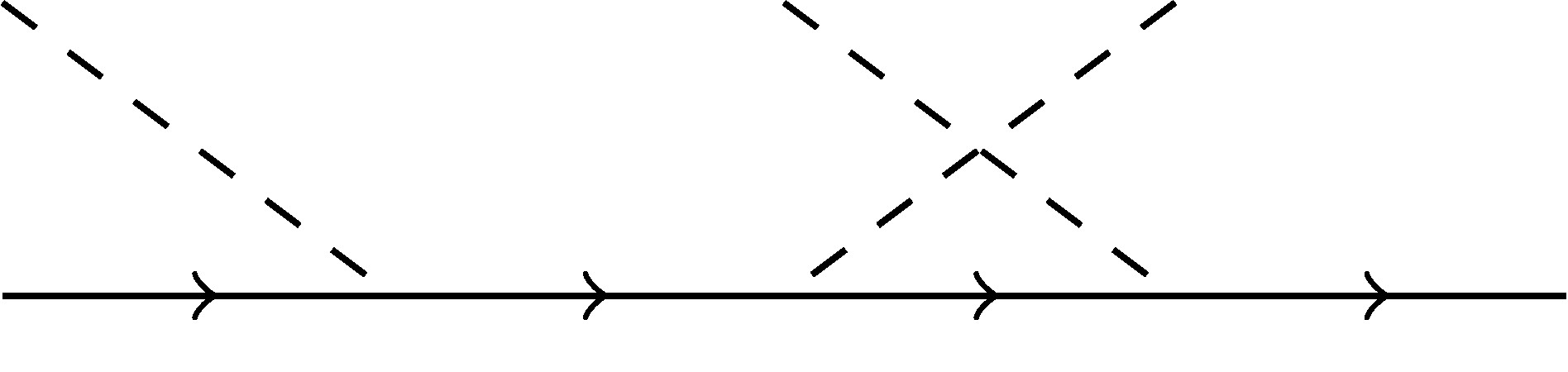}}
+
\raisebox{-.1cm}{\includegraphics{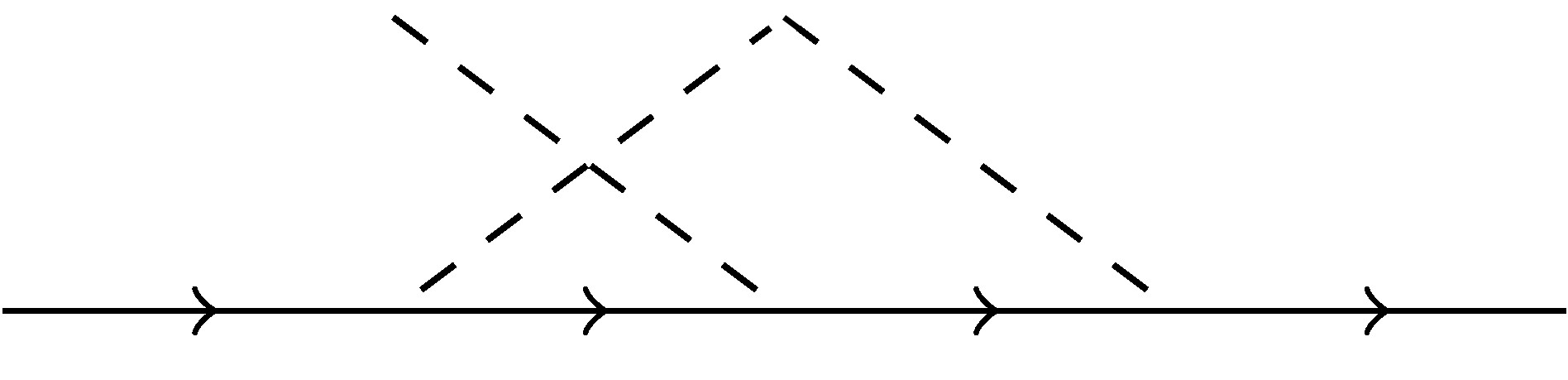}}
	\]
	\caption{Two absorptions and one emission at tree level.}
	\label{Fig:ininout}
\end{figure}

This is an  interesting process as the mixture of absorptions and an emission  will induce both $\laserm$ and $\laserv$ terms, and it is not a priori clear if there  will be  interference between  loop corrections. 
 The tree level process of interest in this respect is thus given by the three processes in Fig.~\ref{Fig:ininout}.

We now expect four sidebands with propagators $\Prop{n+2}$, $\Prop{n+1}$, $\Prop{n}$ and $\Prop{n-1}$. There will also be a mixture of the $v$ term seen in Fig.~\ref{Fig:inandin} and the mass term found in the central band of  Fig.~\ref{Fig:inout}.  The ultraviolet loop corrections will now generate twenty one  graphs. The strategy for evaluating these is again  to group terms so that we get  a mixture of previously evaluated sub-terms and absorption or emission factors analogous to (\ref{eq:ISigmaI}) and (\ref{eq:massid}). The end result of this gives  the loop corrections summarised within the full (tree level and loop)  sideband structures:  
\begin{align}\label{eq:ren_ininout}
\begin{split}
 &
 \big(\tfrac12\In^2-\tfrac12v\big)\Big(\Prop{n-1}+\Prop{n-1}\Sigma_{n-1}\Prop{n-1} \Big)\Out\\
 &
 -\Out\big(\tfrac12\In^2-\tfrac12v\big)\Big(\Prop{n}+\Prop{n}\Sigma_{n+1}\Prop{n} \Big) -\In\Big(\Prop{n}+\Prop{n}\Sigma_{n}\Prop{n} \Big)\Out\In\\
 &\qquad-\In\Big(\Prop{n}+\Prop{n}\Sigma_{n}\Prop{n}\Big)i\big(\mstarsl+\sigmasl\big)\Big(\Prop{n}+\Prop{n}\Sigma_{n}\Prop{n}\Big)\\
 &+\Big(\Prop{n+1}+\Prop{n+1}\Sigma_{n+1}\Prop{n+1} \Big)\Out\big(\tfrac12\In^2+\tfrac12v\big) +\Out\In\Big(\Prop{n+1}+\Prop{n+1}\Sigma_{n+1}\Prop{n+1} \Big)\In\\
 &\qquad+\Big(\Prop{n+1}+\Prop{n+1}\Sigma_{n+1}\Prop{n+1} \Big)i\big(\mstarsl+\sigmasl\big)\Big(\Prop{n+1}+\Prop{n+1}\Sigma_{n+1}\Prop{n+1} \Big)\In\\
&-\Out\Big(\Prop{n+2}+\Prop{n+2}\Sigma_{n+2}\Prop{n+2} \Big)\big(\tfrac12\In^2+\tfrac12v\big)\,.  
\end{split}
\end{align}
Here we see clearly  the same structures in  these loop corrections as encountered earlier in (\ref{eq:ren_in}),
(\ref{eq:ren_out}),  (\ref{eq:ren_inin}), (\ref{eq:ren_outout}) and (\ref{eq:ren_inout}). From this result we can immediately deduce the corresponding dual process involving  two emissions and just one absorption from the background. We see that there is no interference between the mass terms and the $v^*$ insertions.

To summarise, these detailed perturbative investigations show that the loop corrections to the propagation of an electron in a plane wave background  preserve the sideband structures and, through that, induce the expected one loop corrections to the normalisation of the propagators, including the vacuum mass shift.  Unexpectedly, we have seen that the laser induced mass  also has an ultraviolet correction. Having exposed and isolated these loop structures, we now address the (minimal) renormalisation needed for  the extraction of finite, physical results.

\section{Renormalisation} 
 
We have seen, through multiple examples, that the sideband structure of the theory is preserved when loop corrections are included. To understand the renormalisation of the theory, let us consider the  $\ell^{\,\mathrm{th}}$ sideband. For this sideband we have seen that the loop corrections induce a replacement 
\begin{equation}\label{eq:loop_prop}
  \Prop{\ell}\to \Prop{\ell}+\Prop{\ell}\Sigma_\ell\Prop{\ell}\,,
\end{equation}
together with an additional correction to the background induced, mass shift
\begin{equation}\label{eq:loop_lmass}
 -i \mstarsl\to-i (\mstarsl+\sigmasl)\,.
\end{equation}
The ultraviolet divergences in $\Sigma_\ell$ and $\sigmasl$, see (\ref{eq:Sigma_n}) and (\ref{eq:massUV}), signal the need for renormalisation. This we now introduce  by shifting from  bare to renormalised quantities.

It is useful here to  refine our notation, introduced in equation~(\ref{eq:Prop}), for the sideband propagator to include both normal and  induced mass terms, as in (\ref{eq:propmM}).  We thus define
\begin{equation}
  \Prop{\ell}(m,\mstar):=\frac{i}{\psl+\ell\ksl-(m+\mstarsl)+i\epsilon}\,.
\end{equation}
Then the loop corrections, (\ref{eq:loop_prop}) and (\ref{eq:loop_lmass}),  can be written more succinctly  as
\begin{equation}\label{eq:loop_prop_ren}
  \Prop{\ell}(m,\mstar)\to \Prop{\ell}(m,\mstar)+\Prop{\ell}(m,\mstar)\big(\Sigma_\ell-i\sigmasl\big)\Prop{\ell}(m,\mstar)\,.
\end{equation}
To now renormalise this sector of the theory, we follow the usual prescription whereby we first interpret these results as arising from working with the bare Volkov fields and masses: $\psiVbare$, $\mbare$ and $\mstarbare$.  Then we define the physical, renormalised quantities $\psiVphys$, $\mphys$ and $\mstarphys$ by
\begin{equation}\label{eq:Vwf_ren}
  \psiVbare:=\mu^{-\varepsilon}\sqrt{Z_2}\,\psiVphys=\mu^{-\varepsilon}\sqrt{1+\delta_2}\,\psiVphys\,,
\end{equation}
\begin{equation}\label{eq:m_ren}
  \mbare:=Z_m\,\mphys=(1+\delta_m)\,\mphys
\end{equation}
and  
\begin{equation}\label{eq:Vm_ren}
  \mstarbare:=Z_{_{\!\!\mstar}}\,\mstarphys=(1+\deltamstar)\,\mstarphys\,.
\end{equation}  
 These counterterms are then determined by the requirement that when we work with  renormalised quantities, we obtain  finite results.  Note that the mass scale $\mu^{-\varepsilon}$ in the wave function renormalisation factor can be neglected in the leading order analysis presented here.

The full, renormalised sideband propagator  at this order is then, from (\ref{eq:loop_prop_ren}), 
\begin{equation}\label{eq:ren_prop1}
 Z_2^{-1} \Prop{\ell}(\mbare,\mstarbaremin)+\Prop{\ell}(m,\mstar)\big(\Sigma_\ell-i\sigmasl\big)\Prop{\ell}(m,\mstar)\,.
\end{equation}
In the second term of this expression the presence of the loop corrections means that renormalised quantities can be immediately used when working with the leading order  loop corrections. In the first term, though,  we are still explicitly working with the bare fields. 

These  bare quantities can be expanded  to give
\begin{align}
Z_2^{-1} \Prop{\ell}(\mbare,\mstarbaremin)&=(1-\delta_2)\Prop{\ell}\big((1+\delta_m)\,\mphys\,,(1+\deltamstar)\,\mstar\big)\nonumber\\
&= \Prop{\ell}(m,\mstar)+\Prop{\ell}(m,\mstar)\big(-\PropI{\ell}\delta_2-i(m\delta_m+\mstarsl\deltamstar)\big) \Prop{\ell}(m,\mstar)\,.
\end{align}
Thus the renormalised sideband propagator, (\ref{eq:ren_prop1}), becomes
\begin{equation}\label{eq:ren_prop2}
 \Prop{\ell}(m,\mstar)+\Prop{\ell}(m,\mstar)\Sigma_\ell^{\mathrm{R}}\Prop{\ell}(m,\mstar)\,,
\end{equation}
where
\begin{equation}
  \Sigma_\ell^{\mathrm{R}}=\Sigma_\ell-\PropI{\ell}\delta_2-i(m\delta_m+\sigmasl+\mstarsl\deltamstar)\,.
\end{equation}

From this, and equations (\ref{eq:Sigma_n}) and (\ref{eq:massUV}), we see that, independent of the sideband being considered, the minimal renormalisation prescription corresponds to  the familiar results that
\begin{equation}\label{eq:ren1}
    \delta_2=\duv\qquad\mathrm{and}\qquad \delta_m=3\duv\,,
  \end{equation} 
along with the additional requirement that
\begin{equation}\label{eq:ren2}
  \deltamstar=\duv\,,
\end{equation}
where $\duv$ was defined in equation~(\ref{eq:deltaUV}).
Our higher order calculations, in terms of absorptions and emissions,  support the expectation that the renormalisation prescriptions (\ref{eq:ren1}) and (\ref{eq:ren2}) holds also in the strong field sector. We will now recall how the sideband formulation can be extended to all such orders and, through this,     conjecture the form of the one-loop corrections to the full Volkov description of an electron propagating through a plane wave, laser background.

\section{The full Volkov description at one loop} 
The importance  of the Volkov solution for the tree level results is that the sideband description, discussed above in the perturbative framework, is known to all orders in the background interaction for this wide class of polarisations, see~\cite{Lavelle:2017dzx}. We now want to develop the  link between the perturbative loop structures presented  here  and that all orders formalism. In doing so we shall see that the perturbative results actually motivate a significant  simplification to the all orders  description. Armed with that result, we shall be able to conjecture a compact expression for the leading one-loop corrections at all orders in the intensity of the background. 

The exact tree level solution for the two point function describing an electron propagating in an elliptically polarised background can be written (see equation (44) in ~\cite{Lavelle:2017dzx} and discussions therein) as the usual momentum space integration factors times the  double sum over $r$ and $s$ of the following  sideband structures: 
\begin{align}\label{eq:two_pnt_volkov}
\begin{split}
  \ee^{ir\kx} \Big(\J_{s+r}(p)&+\frac{\ksl\lasersl}{2\pk}\J_{s+r+1}(p)+\frac{\ksl\laserssl}{2\pk}\J_{s+r-1}(p)\Big)\\&\qquad\quad\times
 \Prop{s}(m,\mstar)\Big(\Js_{s}(p)-\frac{\ksl\laserssl}{2\pk}\Js_{s+1}(p)-\frac{\ksl\lasersl}{2\pk}\Js_{s-1}(p)\Big)\,.   
  \end{split}
\end{align}

Unpicking (\ref{eq:two_pnt_volkov}) we see that, as we sum over $s$, the sideband propagator $\Prop{s}(m,\mstar)$ is sandwiched between  factors built out of  (generalised) Bessel functions, $\J_\ell(p)$, where the parameter $\ell$  can be various  combinations of the summation parameters $r$ and $s$. These Bessel functions are also labelled by the eccentricity parameter $\tau$ characterising the polarisation of the background in the elliptic class. The precise definition of these Bessel functions is that
\begin{equation}\label{eq:genBess}
 \J_\ell(p):= J_\ell(\omega_1,v,\omega_2)=\frac1{2\pi}\int_{-\pi}^\pi \!\!d\theta \,
  \ee^{i(\omega_1\sin\theta+v\sin2\theta+\omega_2\cos\theta)}\ee^{-i\ell\theta}\,,
\end{equation}
where the  eccentricity information is now encoded in the real parameters $\omega_1$, $v$ and $\omega_2$. The connection with the  complex vector  parameters  $\laser$ and $\lasers$,  introduced in (\ref{eq:Ais})  and (\ref{eq:Eis}),  is seen in equation (\ref{eq:vdef}) for $v$, and the definitions
\begin{equation}
  \omega_1=-\left(\frac{\plaser}{\pk}+\frac{\plasers}{\pk}\right)
  \qquad \mathrm{and}\qquad
  \omega_2=-i\left(\frac{\plaser}{\pk}-\frac{\plasers}{\pk}\right)\,.
\end{equation}
The fact that $v$ is real is perhaps surprising and seems at odds with our effort, as in (\ref{eq:vsdef}), to distinguish typographically between $v$ and $v^*$. However, this was a useful bookkeeping  device to keep track of the duality structures seen earlier, and one that we will return to below.

From the perturbative perspective, one of the most striking and immediate things to note  about the all orders result (\ref{eq:two_pnt_volkov}) is the absence of the variables that were the building blocks in the description developed in this paper. In particular,  the In and Out terms, (\ref{eq:In}) and (\ref{eq:Out}), seem to be absent. 

Give the central role played by these terms in our perturbative analysis, it seems logical to try to rewrite the all orders result in terms of them.  To  this end,  we make the change of variables  $\omega_1\to\Omega_1$ and $\omega_2\to\Omega_2$, with
\begin{equation}\label{eq:Omeg1IO}
  \Omega_1=\omega_1-\frac{\ksl\lasersl-\ksl\laserssl}{2\pk}=-(\In+\Out)
\end{equation} 
and
\begin{equation}\label{eq:Omeg2IO}
  \Omega_2=\omega_2-i\frac{\ksl\lasersl+\ksl\laserssl}{2\pk}=-i(\In-\Out)\,.
\end{equation}
Note that the reality requirements on $\omega_1$ and $\omega_2$ are now replaced by the duality result that $\overbar{\Omega}_1=\Omega_1$ and $\overbar{\Omega}_2=\Omega_2$. The trivial commutativity of $\omega_1$ and $\omega_2$ is now the non-trivial matrix result that $\Omega_1\Omega_2=\Omega_2\Omega_1$, which is ensured by the null properties of the background field.

These commutativity and duality relations enable us to extend the domain of the  Bessel functions defined in~(\ref{eq:genBess}) so that we can unambiguously write
\begin{equation}\label{eq:genBessOmega}
 J_\ell(\Omega_1,v,\Omega_2):=\frac1{2\pi}\int_{-\pi}^\pi \!\!d\theta \,
  \ee^{i(\Omega_1\sin\theta+v\sin2\theta+\Omega_2\cos\theta)}\ee^{-i\ell\theta}\,.
\end{equation}
To understand the connection between these extended functions and the complicated pre and post factors in the two point function  (\ref{eq:two_pnt_volkov}), we note that the null property of the vector  $k$ means that
\begin{equation}
   \ee^{i\Omega_1\sin\theta}=\ee^{i\omega_1\sin\theta}\left(1-i\frac{\ksl\lasersl-\ksl\laserssl}{2\pk}\sin\theta\right)
 \end{equation} 
 and
\begin{equation}
   \ee^{i\Omega_2\cos\theta}=\ee^{i\omega_2\cos\theta}\left(1+\frac{\ksl\lasersl+\ksl\laserssl}{2\pk}\cos\theta\right)\,.
 \end{equation}
 Hence we quickly see that
 \begin{equation}
   J_\ell(\Omega_1,v,\Omega_2)=J_\ell(\omega_1,v,\omega_2)+J_{\ell+1}(\omega_1,v,\omega_2)\frac{\ksl\lasersl}{2\pk}+J_{\ell-1}(\omega_1,v,\omega_2)\frac{\ksl\laserssl}{2\pk}
 \end{equation}
 and
\begin{equation}
   \overbar{J}_\ell(\Omega_1,v,\Omega_2)=J_\ell^*(\omega_1,v,\omega_2)-J_{\ell+1}^*(\omega_1,v,\omega_2)\frac{\ksl\laserssl}{2\pk}-J_{\ell-1}^*(\omega_1,v,\omega_2)\frac{\ksl\lasersl}{2\pk}\,.
\end{equation}
We can thus write the two point function  (\ref{eq:two_pnt_volkov}) in a much more compact way as  the sum of all terms of the form
\begin{equation}\label{eq:propOmega}
\ee^{ir\kx}  J_{s+r}(\Omega_1,v,\Omega_2)\Prop{s}(m,\mstar)\overbar{J}_{s}(\Omega_1,v,\Omega_2)\,.
\end{equation} 

To link this with our perturbative results, it is instructive to consider the  $r=-1$ terms in this double sum  with $s$ ranging from -1 to 2. Expanding the Bessel functions (\ref{eq:genBessOmega}) in terms of the In and Out representations, (\ref{eq:Omeg1IO}) and (\ref{eq:Omeg2IO}),   gives for  this part of (\ref{eq:propOmega}) the explicit sum:
\begin{align}\label{eq:rminusone}
  \begin{split}
    \ee^{-i\kx}\bigg(&\big(\tfrac12\In^2-\tfrac12v\big)\Prop{-1}(m,\mstar)\Out\\
    &-\Out\big(\tfrac12\In^2-\tfrac12v\big)\Prop{0}(m,\mstar)+\In \Prop{0}(m,\mstar)\big(1-\In\Out\big)\\
    &\quad+\Prop{1}(m,\mstar)\Out\big(\tfrac12\In^2+\tfrac12v\big)-\big(1-\In\Out\big)\Prop{1}(m,\mstar)\In\\
    &\qquad-\Out\Prop{2}(m,\mstar)\big(\tfrac12\In^2+\tfrac12v\big)\bigg)\,.
  \end{split}
\end{align}
These terms are precisely the sum of the sidebands  derived in (\ref{eq:Sb1}) and the tree level part of (\ref{eq:ren_ininout}), both with $n=0$.  

This identification of the momentum dependence in (\ref{eq:rminusone}) with perturbative structures is gratifying and hints at the underlying logic of how to group the perturbative terms together.

In the perturbative formulation developed here we have not yet incorporated the fact that the laser background breaks translational invariance. This means that our  momentum space description, where we have presented a direct way to calculate loop corrections for the sidebands, requires modification. 

From the exact solution (\ref{eq:two_pnt_volkov}) we see that the modification is very simple. In addition to the standard momentum space factor $e^{-ip\cdot(x-y)}$ which multiplies~(\ref{eq:two_pnt_volkov}), we see an $e^{ir\kdotx}$ factor which explicitly violates translation invariance. This, though, can be exploited to organise the perturbative discussion and will allow us to group terms consistently.

The key observation to note is that all the terms in (\ref{eq:rminusone}) share a common  homogeneity  in the absorption and emission fields. Indeed, they  all include  either an absorption and no emissions, or two absorptions and one emission. From this simple observation it follows that if we multiply each absorption term by a factor of $e^{-i\kdotx}$ and each emission term by $e^{i\kdotx}$, then we obtain the overall phase factor seen in~(\ref{eq:rminusone}). In a similar way, the terms in Fig.~\ref{Fig:inandin}, say, would be accompanied by a factor of $e^{-i2\kdotx}$, while  Fig.~\ref{Fig:outandout} would pick up a factor of $e^{i2\kdotx}$.

This motivates the following redefinition of the fundamental absorption vertex~(\ref{eq:Ais}) by including the exponential factor: 
\begin{equation}
     \Ab=-i\,\lasersl\to -i\,\ee^{-i\kx}\lasersl\,.
\end{equation} 
Similarly, we have the associated dual redefinition $\Em:=- \overbar{\Ab}\to-i\,\ee^{i\kx}\laserssl$.  Hence from~(\ref{eq:vdef}) we see that  $v\to \ee^{-i2\kx}v$ while, from~(\ref{eq:vsdef}), $v^*\to \ee^{i2\kx}v^{*}$. In terms of these redefinitions, $v$ and $v^*$ are not the same, so the notational convenience used earlier now becomes a genuine  distinction. It is also clear now how to combine the perturbative terms in a physically correct  manner in terms of commensurate powers of absorption minus emission. Note that the mass term $\mstarsl$ picks up no spatial dependence under this redefinition.

\begin{figure}[htb]
	\[
\raisebox{-0.5cm}{\includegraphics{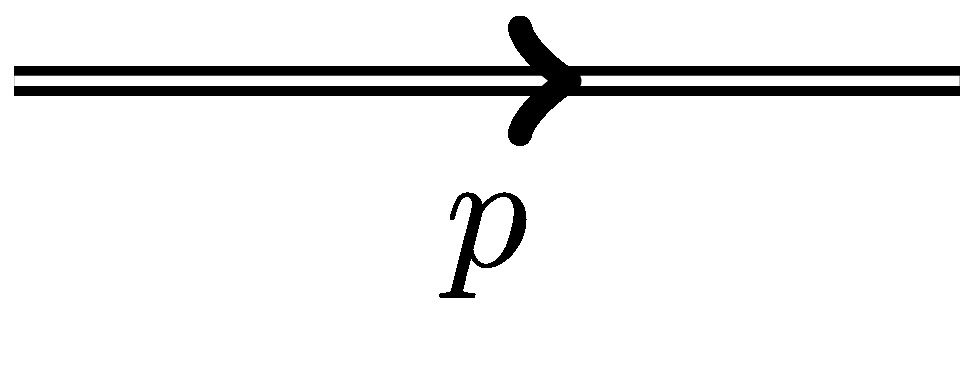}}+\raisebox{-0.6cm}{\includegraphics{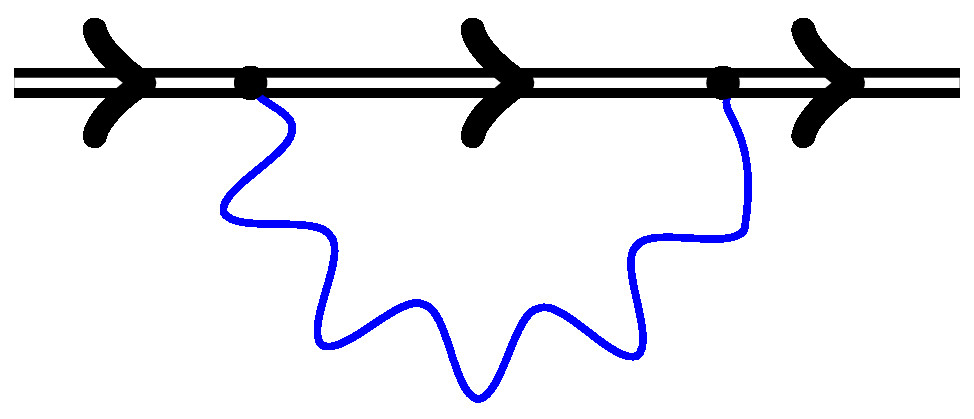}}= \sum_r\sum_sJ_{s+r}(\In:v:\Out)\Prop{s}(m,\mstar)\overbar{J}_{s}(\In:v:\Out)
	\]
	\caption{Strong field one loop conjecture.}
	\label{Fig:conjecture }
\end{figure}

Armed with this reformulation of the theory, we  now  conjecture  how the double line description Fig.~\ref{fig:double_line} and its one loop corrections symbolised by Fig.~\ref{fig:double_line_loop} should be defined in terms of the renormalised fields (\ref{eq:Vwf_ren}), (\ref{eq:m_ren}) and (\ref{eq:Vm_ren}), within  the minimal subtraction scheme defined by (\ref{eq:ren1}) and (\ref{eq:ren2}). Our conjecture is that, in terms of the renormalised masses introduced in this paper, we have the  identification summarised in Fig.~\ref{Fig:conjecture },
where we have introduced the condensed notation that  $J_s(\In:v:\Out)=J_s(-(\In+\Out),v,-i(\In-\Out))$. 

This result holds at the tree level to all orders, and we have seen in this paper that, at one loop in Feynman gauge,  it also holds for the ultraviolet poles in several different sidebands.
\section{Conclusions}

In this paper we have  developed a  perturbative description of the propagation of an electron in a plane wave background. There are two expansions here: one in the interactions with the background and an expansion up to one loop in perturbation theory. Each interaction with the background generates sideband structures. We have seen that the loop corrections maintain these sidebands. This means that a multiplicative renormalisation of the theory can be carried out in this formulation. We have worked in Feynman gauge and the background chosen was the full elliptic class of polarisations to bring out any polarisation dependence.

The tree level sideband approach to charge propagation in a laser has the advantage that, at its heart, it identifies with each sideband a standard propagator, with momentum shifted by some multiple of the background momentum. These propagators are then multiplied by well defined terms characterising the laser. We have carried out a weak field expansion and explicitly calculated leading contributions to various sidebands.

Using dimensional regularisation, we have calculated the one loop, ultraviolet divergent poles in these sideband structures. They included multiple absorptions, multiple emissions, and, importantly, contributions from a mixture of absorptions and emissions.  These last structures are responsible for the background induced electron mass shift. 

Our calculations have revealed that the loop corrections to the sidebands replace the propagators by their equivalent standard one loop corrections. This is a minimal requirement for multiplicative renormalisation. However, we also found one additional ultraviolet divergent correction. This pole is a correction to the laser induced mass shift. As this was unexpected, we have verified that the same correction occurs in different sidebands in the Volkov propagator.   

We have seen how to renormalise these divergences in terms of the usual one loop renormalisation without a background, plus an additional multiplicative renormalisation of the laser induced mass shift. Inspired by the all orders tree level description, we have  been able to conjecture an all orders expression for the full one loop corrections in this class of backgrounds.  

To complete this conjecture for the pole structure requires a proof that it holds to all orders in emissions and absorptions from the laser. A stronger form of the conjecture involves showing that the ultraviolet finite loop corrections, and any infrared divergences~\cite{Lavelle:2005bt}, are also compatible with this structure. Finally, it is important to study these results in other gauges.   

\section{Acknowledgements} 
We thank Tom Heinzl, Anton Ilderton and Ben King for discussions and comments.

%
%
%
%

\end{document}